\documentstyle [12pt,epsfig]{article} 
\makeatletter
\@addtoreset{equation}{section}
\makeatother

\newcommand{\ba}{\begin{eqnarray}}
\newcommand{\ea}{\end{eqnarray}}

\begin{document}
\newcommand{\BS}{\bigskip}
\newcommand{\SECTION}[1]{\BS{\large\section{\bf #1}}}
\newcommand{\SUBSECTION}[1]{\BS{\large\subsection{\bf #1}}}
\newcommand{\SUBSUBSECTION}[1]{\BS{\large\subsubsection{\bf #1}}}

\begin{titlepage}
\hspace*{8cm} {UGVA-DPNC 1998/04-176 April 1998}
\begin{center}
\vspace*{2cm}
{\large \bf
SPACE TIME MEASUREMENTS IN SPECIAL RELATIVITY}  
\vspace*{1.5cm}
\end{center}
\begin{center}
{\bf J.H.Field }
\end{center}
\begin{center}
{ 
D\'{e}partement de Physique Nucl\'{e}aire et Corpusculaire
 Universit\'{e} de Gen\`{e}ve . 24, quai Ernest-Ansermet
 CH-1211 Gen\`{e}ve 4.
}
\end{center}
\vspace*{2cm}
\begin{abstract}
 The conventional discussion of apparent distortions of space and time in
 Special Relativity (the Lorentz-Fitzgerald Contraction and Time Dilatation)
 is extended by considering observations of : (i) moving objects of limited
 lifetime in their own rest frame (`transient luminous objects') and (ii)
 a moving extended array of synchronised `equivalent clocks' in a common
 inertial frame. Application of the Lorentz Transformation to (i) shows
 that such objects, viewed with coarse time resolution, appear to be 
 {\it longer} in the direction of the relative velocity $\vec{v}$ by 
 a factor $1/\sqrt{1-(v/c)^2}$ (Space Dilatation) and to (ii) that the
 moving equivalent clock that appears at any fixed position in the
 rest frame of
 an inertial observer appears to be running {\it faster} than a similar
 clock at rest by the factor $1/\sqrt{1-(v/c)^2}$ (Time Contraction).
 The actual appearance of moving objects and clocks, taking into account 
 light propagation time delays, as well as the effect of the Lorentz 
 Transformation, is also discussed.
\end{abstract}
\vspace*{4cm}
\vspace*{1cm}
{\it Published in the Proceedings of the XX Workshop on High Energy
Physics and Field Theory, Protvino, Russia, June 24-26 1997.} 
\end{titlepage}

\SECTION{\bf{Introduction}}
 In his 1905 paper on Special Relativity~\cite{x1}Einstein showed that
  Time Dilatation (TD) and the Lorentz-Fitzgerald Contraction (LFC), which had
  previously been introduced in a somewhat ad hoc way into Classical 
  Electrodynamics, are simple consequences of the Lorentz Transformation (LT),
  that is, of the geometry of space-time. 
  \par As an example of the LFC Einstein stated that a sphere moving with
   velocity $v$ would, `viewed from the stationary system', appear to be 
   contracted by the factor $\sqrt{1-(\frac{v}{c})^2}$ in its direction of
    motion where $c$ is the velocity of light in free space. It was only pointed
    out some 54 years later that if `viewed' was interpreted in the conventional
    sense of `as seen by the eye, or recorded on a photograph' then the sphere 
 does not at all appear to be contracted, but is still seen as a sphere with 
 the same dimensions as a stationary one and at the same position~\cite{x2,x3,x4} !
 It was shown in general~\cite{x3,x4} that transversely viewed moving objects
 subtending a small solid angle at the observer appear to be not distorted in shape
 or changed in size, but rather rotated, as compared to a similarly viewed and 
 orientated object at rest. This apparent rotation is a consequence of three distinct
 physical effects:
 \begin{itemize}
 \item[(i)] The LFC.
 \item[(ii)] Optical Aberration.   
 \item[(iii)] Different propagation times of photons emitted by different
 parts of the moving object.   
 \end{itemize}
 The effect (ii) may be interpreted as the change in direction of photons, emitted 
 by a moving source, due to the LT between the rest frames of the source and the
 stationary observer. Correcting for (ii) and (iii), the LFC can be deduced as a 
 physical effect, if not directly observed. It was also pointed out by Weinstein
 ~\cite{x5} that if a single observer is close to a moving object then, because of
 the effect of light propagation time delays it will appear elongated if moving 
 towards the observer and contracted (to an extent greater than the LFC) if moving 
 away. Only an object moving strictly transversely to the line of sight of a 
 close observer shows the LFC.
 \par However, the LFC itself is a physical phenomenon similar in many ways to (iii)
 above. A photograph or the human eye record as a sharp image the photons incident on
 it at a fixed time. That is, the image corresponds to a projection at constant time
 in the frame S of observation. This implies that the photons constituting the image
 are emitted at different times from the different parts, along the line of sight, of
 an extended object. The LFC is similarly defined by a fixed time projection in the
 frame S. The LT then requires that the photons constituting the image of a
 moving object
 are also emitted at different times, in
 the rest frame S' of the object, from the different parts along
 its direction of motion. In the following S will, in general, denote the
 reference frame of a `stationary' observer (space-time coordinates x,y,z,t)
 while S' refers to the rest frame of an object moving with uniform
 velocity $v$ relative to S ( space-time coordinates x',y',z',t').  
 \par The purpose of this paper is to point out that the $t = $ constant projection
 of the LFC and the $x' = $ constant projection of TD are not the only physically
 distinct Space Time Measurements (STM) possible within Special Relativity.
 In fact, as will be demonstrated below, there are two others: Space Dilatation (SD),
 the $t' = $ constant projection and Time Contraction (TC), the $x = $ constant 
 projection. The overall situation with respect to apparent distortions of space-time
 is thus symmetric with respect to space and time. The $t = $ constant projection of 
 the LFC is the STM appropriate to the `moving bodies' of Einstein's original paper
 and to the photographic recording technique. This medium has no intrinsic time 
 resolution and relies on that provided by a rapidly moving shutter to provide
 a clear image. The LFC `works' as a well defined physical phenomenon
 because the `measuring rod' or other physical object under observation has
 a lifetime that is long in comparison with the time interval required to make an
 observation, and so constitutes a continuous source of emitted or reflected 
 photons, such that some are always available in the different space ($\Delta x'$)
 and time ($\Delta t'$) intervals in S' for every position of the rod 
 corresponding to the time interval $\Delta t$ around $t = $ constant in the 
 observer's frame S. If, however, the physical object of interest has internal motion
 (rotation, expansion or contraction) or is only illuminated, in its rest frame S', 
 during a short time interval, the above conditions, that assure that the 
 $t = $ constant projection gives a well defined STM no longer apply. In the following,
 for brevity, all such objects of limited luminous lifetime in their own rest frame
 will be referred to as `Transient Luminous Objects' or TLO. For such objects it is
 natural to define a length measurement by taking the $t' = $ constant projection
 in S'.
 \par Space time measurements of such transient luminous objects are discussed 
 below in Section 3. A simple conceptual camera of finite time resolution is considered.
 Practical examples are a flash camera or a TV raster. The camera is used to record 
 Image Plane Coordinates (IPC) $x_I,~y_I,~t_I$ in the observation frame S. To orient the
 discussion answers will be sought to the question: `What information about transient
 luminous objects (~observed in general as a correlated ensemble of STM in the frame S)
 such as its distance, shape, orientation or lifetime can be derived {\it only} from
 measurements of the image plane coordinates? The case of stationary objects is first
 considered in Section 2, followed by the discussion of uniformly (transversely) 
  moving objects in the following Section.
  \par In Section 4 time measurements other than the conventional TD 
  ($x' = $ constant projection) of Special Relativity are considered. The TD phenomenon
  refers only to a local clock, in the sense that its position in the frame S' is
  invariant (say at the spatial origin of coordinates). However the time recorded by
  any synchronised clock in the same inertial frame is, by definition, identical.
  Einstein used such an array of `equivalent clocks' situated at different positions
  in the same inertial frame in his original discussion of the relativity of 
  simultanaeity~\cite{x1}. The question addressed in Section 4 is: 
  What will an observer in S see if he looks not only at a given local clock in S',
  but also at other equivalent clocks at different positions in S', in comparison
  to a standard clock at rest in his own frame? It is shown that such equivalent clocks
  may be seen to run slower than, or faster than, the TD prediction for a local clock.
  In particular they may even appear to {\it run faster than the standard clock}. 
  This is an example of the Time Contraction effect mentioned above.   
 \SECTION{\bf{Space Time Measurements of Transient Luminous Objects at Rest}}
It is part of common experience that distant objects appear smaller than similar ones that 
are close to the observer. This is because the human visual system functions as a camera.
In Fig 1 is shown a schematic camera where a number of planar objects $O_A$,$O_B$,$O_C$,$O_D$
of different shape, orientation and distance produce an identical image (a square) in the image
plane of the camera. The lens of the latter will be considered in the following discussion 
to be a simple pin-hole. If the size and orientation of the object is known then its distance
$l$ from the position $L$ of the camera lens along the z-axis may be deduced from measurements
of its image by simple geometry. For example a rod of known length $r$, whose centre lies
on the z-axis, and is orientated parallel  to $Ox_I$ is at the distance 
$l = rf/r_I$ where $r_I$ is the length of the image of the rod and $f$ is 
the distance between the lens and the image plane of the camera.
\begin{figure}[htbp]
\begin{center}\hspace*{-0.5cm}\mbox{
\epsfysize10.0cm\epsffile{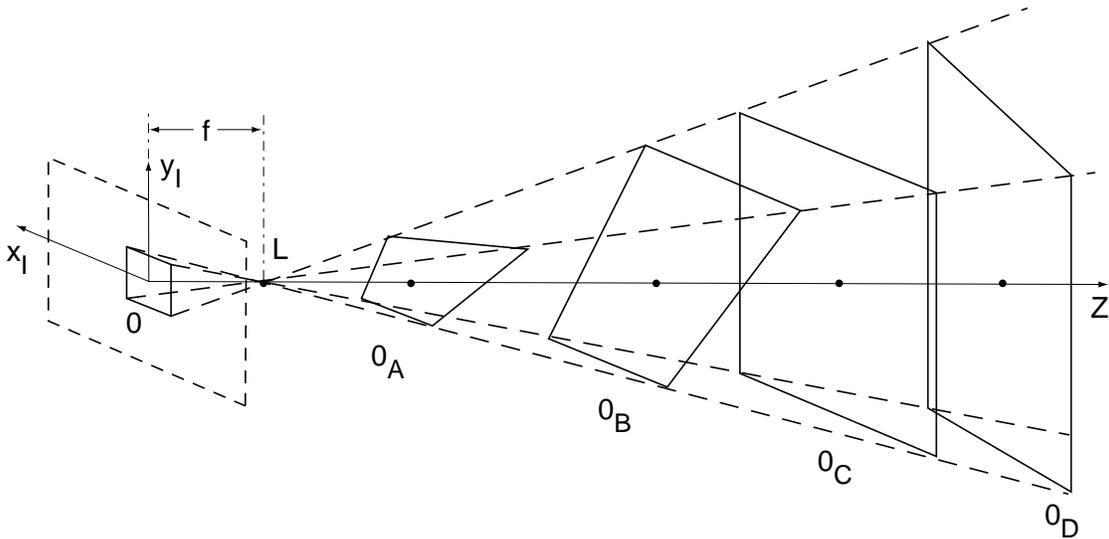}}
\caption{ The schematic camera and the four planar objects $O_A$, $O_B$, $O_C$, $O_D$.  }
\label{fig-fig1}
\end{center}
 \end{figure}  
\begin{figure}[htbp]
\begin{center}\hspace*{-0.5cm}\mbox{
\epsfysize10.0cm\epsffile{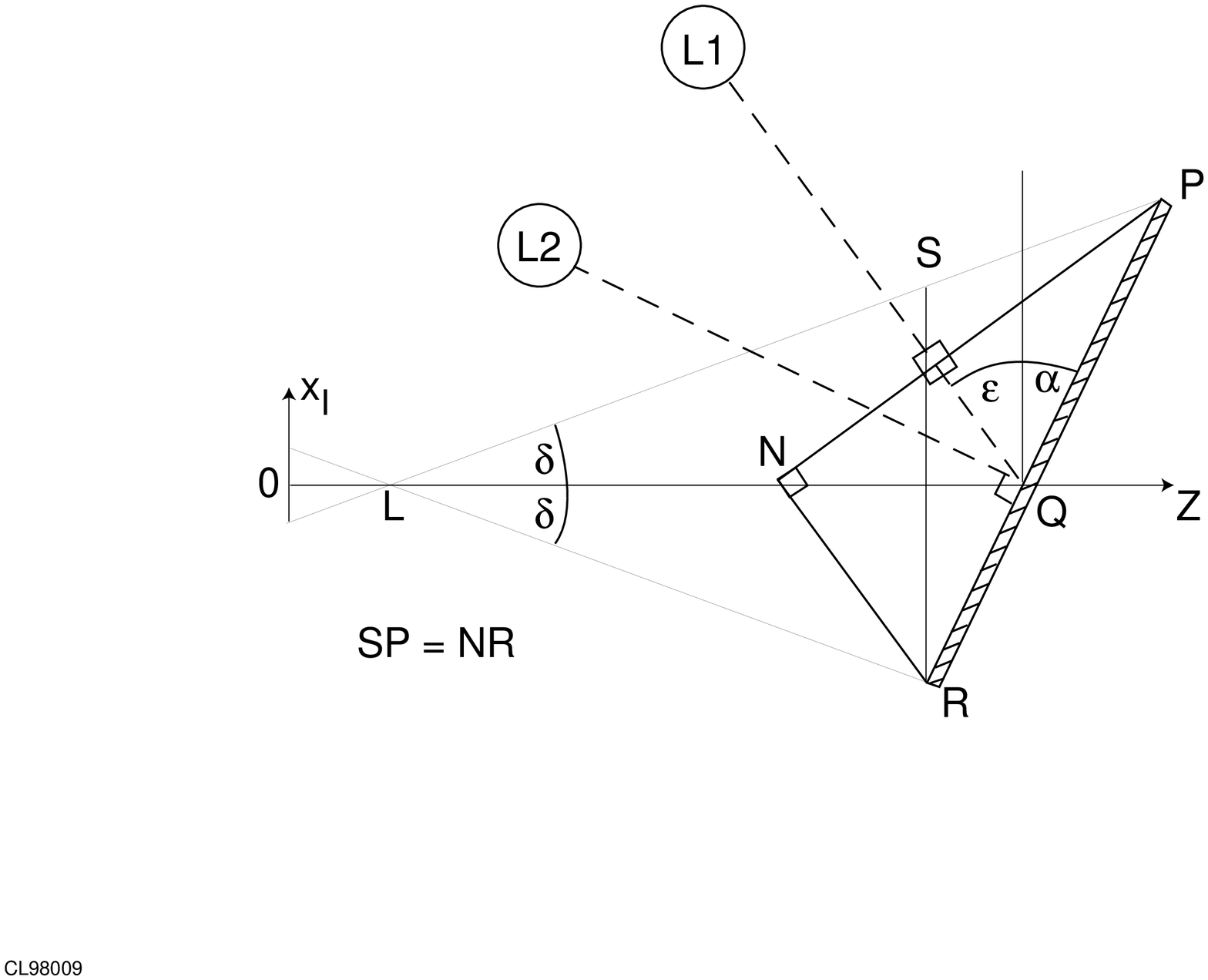}}
\caption{ The projection into the $x_I$-$z$ plane of the object $O_D$, showing
the positions in this plane of the lamps L1 and L2.}
\label{fig-fig2}
\end{center}
 \end{figure}
\begin{figure}[htbp]
\begin{center}\hspace*{-0.5cm}\mbox{
\epsfysize10.0cm\epsffile{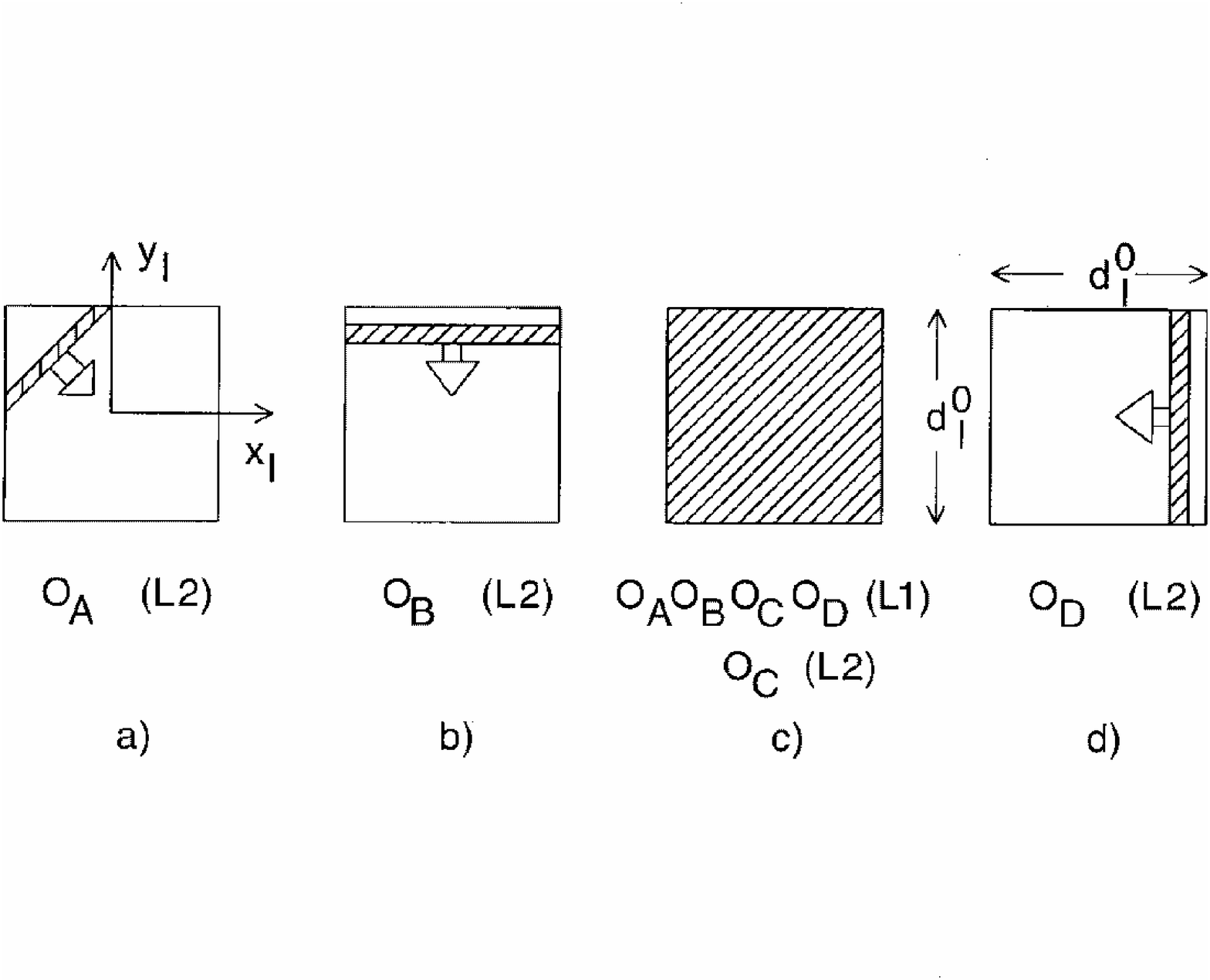}}
\caption{Images of the objects observed in the camera
 (viewed from the side of the objects) when lamps L1 and L2 are flashed. The arrows show the direction 
of motion of the instantaneous line image. The squares show the total areas
swept out in the image plane by the line image. a),b),c),d) correspond to
 $O_A$, $O_B$, $O_C$ and $O_D$ for L2 whereas L1 gives the image shown in c) for all four
 objects.} 
\label{fig-fig3}
\end{center}
 \end{figure}
\par The problem addressed now is how the real dimensions and positions of a physical
object may, in general, be deduced only from measurements made in the image plane of the camera
shown in Fig 1. One possiblity is to send a pulse of light with a sharp time distribution
at a known time from a source near the camera and to measure the time delay and the positions of
the photons scattered back into the camera. This is the principle of Radar ranging 
measurements. Here a different, although related, method will be used where the camera
requires no external time reference. In the present Section the method will be applied 
to objects at rest relative to the camera, while in the following Section it is applied to 
objects in uniform transverse motion with respect to the camera axis. In this case effects
of Special Relativity must be taken into account.
\par Two different lamps L1, L2 are used to illuminate the object of interest with a
sharp pulse of light whose width must be smaller than $d/c$ where $d$ is the typical
transverse dimension of the object. The lamps L1, L2 are used as a source of correlated
space-time events in the rest frame of the object of interest, that is to produce a TLO
of well defined properties. They are not however essential for the following considerations.
The surfaces of the objects could just as well be equipped with an array of discrete light
sources (e.g. light-emitting diodes) programmed so as to emit light pulses at different values
of $t'$, the proper time of the object, depending on their position.
\par The lamp L1 is situated in such a position that the photons scattered from different
parts of the surface arrive simulaneously at the image plane of the camera
\footnote{This is true in the limit that $\delta^2$ is negligible as compared to
unity}. That is the position of the lamp is such that the `trivial' light propagation
time differences from different parts of the object to the camera are compensated.
If L1 lies in the plane spanned by the z-axis and a normal to the surface of the object, then 
the angle $\varepsilon$ (Fig 2) is given by:
\begin{equation}
\varepsilon = \cos^{-1}(\frac{\sin \alpha}{\cos \delta})-\alpha
\end{equation}
The lamp L2 is placed along the normal to the surface of the object on the same side as the
camera. Thus the scattering time of the signal from L2 corresponds to a fixed proper
time in the rest frame of the (planar) object. In order that the light from the lamps
may enter the camera in the absence of specular reflection the surfaces of the 
objects should be diffuse reflectors. If the lamps are sufficiently distant the impinging
light pulse can be considered to be a narrow planar wave packet.
\par For each of the objects $O_A$-$O_D$ the light pulse from L1 gives a sharp (in time) square
image in the camera (Fig 3c). No information is therefore available as to which of the four objects
in Fig 1 produced the image. For the lamp L2, because of light propagation time differences,
each object produces a clearly distinguishable image, a line moving in a distinct direction,
the extremities of which sweep out the same square outline (Figs 3a,b,d) or, for $O_C$, an
instantaneous square image indistinguishable
\footnote{On the assumption that the transverse dimensions of the object are much larger than 
those of the camera and that L2 is more distant from the object than the camera, the latter will,
for $O_C$, shadow the central region of the square image. This inessential complication does not
arise if $O_C$ is self luminous.}
 from that produced by the lamp L1 (Fig 3c).
The orientation of the moving line image defines that of the surface of the object.
It is perpendicular to the projection of the normal to the surface on to the image plane.
It is now demonstrated that analysis of the time dependant images of $O_A$, $O_B$, $O_D$ enables
each object to be spatially reconstructed. Allowing for light propagation delays, the ensemble of 
space-time points constituting the TLO may then also be calculated.
\par As an example the image produced by $O_D$ using the lamp L2 is now analysed. The times for the
line to move from the right extremity of the square to the centre, and from the centre to the left
extremity are denoted by $\Delta t_1$, $\Delta t_2$ respectively. With $LQ = l$, and other 
geometrical definitions as in Fig 4a, then: 
\begin{eqnarray}  
\Delta t_1^{L2} & = & \frac{(2ld_1 \tan \alpha - d_1^2 \sec^2 \alpha)}
{c[l(1+\cos \delta)-d_1 \tan \alpha ]} \cos \delta \\
\Delta t_2^{L2} & = & \frac{(2ld_2 \tan \alpha + d_2^2 \sec^2 \alpha)}
{c[l(1+\cos \delta)+d_2 \tan \alpha ]} \cos \delta
\end{eqnarray}
\begin{equation}
\frac{d_1}{l-d_1 \tan \alpha} =  \frac{d_2}{l+d_2 \tan \alpha} = \tan \delta
\end{equation}
In Eqns(2.2),(2.3) light propagation time differences inside the camera are 
neglected. With the further assumptions $\delta \ll 1$,~ $ l \gg \frac{d_2}{2}
 \tan \alpha,~ d_2 \rm{cosec} 2 \alpha$~ Eqns(2.2),(2.3) simplify to :
\begin{eqnarray}
\Delta t_1^{L2} & \simeq & \frac{d_1}{c} \tan \alpha \\
\Delta t_2^{L2} & \simeq & \frac{d_2}{c} \tan \alpha
\end{eqnarray}
Denoting by $d_I^0$ the size of the side of the square swept out by the
moving line image in Fig 3d then:
\begin{equation}
\frac{d_I^0}{2f} = \tan \delta
\end{equation}
Eqns(2.4-2.7) have the following solution:
\begin{eqnarray}
d_1 & = & \frac{d_I^0}{2f} c \Delta t_1 A_t^{L2}  \\
d_2 & = & \frac{d_I^0}{2f} c \Delta t_2 A_t^{L2}  \\ 
\cot \alpha & = & \frac{d_I^0}{2f} A_t^{L2}  \\
l & = & \frac{2c\Delta t_1^{L2} \Delta t_2^{L2}}{\Delta t_2^{L2}-\Delta t_1^{L2}}
\end{eqnarray}
where:
\[ A_t^{L2} = \frac {\Delta t_1^{L2}+ \Delta t_2^{L2}}{\Delta t_2^{L2}-\Delta t_1^{L2}} \]
Thus the position, orientation and physical dimensions of $O_D$ are completely specified
by measurements of $\Delta t_1^{L2}$, $\Delta t_2^{L2}$ and $d_I^0$, and the orientation 
of the moving line image. Similar calculations may be performed, {\it mutandis mutandi} for 
$O_A$ and $O_B$.
The object $O_C$ is a special case. Since $ \Delta t_1^{L2} = \Delta t_2^{L2} = 0$
it must be orientated perpendicular to Oz. However no information can be obtained, using the 
lamp L2, about its position and size.

 \SECTION{\bf{Space Time Measurements of Transient Luminous Objects in Uniform
 Transverse Motion}}
The observation of the object $O_D$ described in the previous Section is is now repeated
in the case that $O_D$ and the lamps L1,L2 are moving with a constant velocity $v =\beta c$
relative to the camera and parallel to O$x_I$.
A coordinate system is defined in the moving frame S' with origin at Q and axes Q$x'$, Q$y'$,
Q$z'$ parallel to O$x_I$, O$y_I$, O$z$ (Fig 4). The frame, at rest relative to the 
camera, whose origin coincides with Q at $t=t'=0$, and with $x$ and $y$ axes parallel to
O$x_I$ and O$y_I$ respectively is denoted by S. It is further assumed that the light pulses
from L1 and L2 arrive at Q at time $t'=0$. To calculate the moving image observed in the 
camera under these conditions the light propagation times found in the previous Section
may be used after first performing the LT of space-time points beween the frames S and S':
\begin{eqnarray}
x & = & \gamma (x'+vt') \\
t & = & \gamma (t'+\beta \frac{x'}{c})
\end{eqnarray}
where
\[ \gamma \equiv \frac{1}{\sqrt{1-\beta^2}} \]
The results of the LT for the points P,Q,R of $O_D$ are summarised in Tables 1 and 2,
for observations using lamps L1 and L2 respectively, under the conditions described above.
The fourth column of Table 2 is an example of the following general result 
(Space Dilatation) referred to in the introduction:
\par {\it A transient luminous object, lying along the Ox' axis, whose length in S' at
fixed time t' is d, appears if observed in S with coarse time resolution to be of 
length $\gamma d$.}
\begin{table}
\begin{center}
\begin{tabular}{|c|c c c c|} \hline
Point on object &  x' &  t' &  x  & t  \\ \hline
   &  &  &  &  \\
P &  $d_2$ & $-\frac{d_2}{c} \tan \alpha$  &
  $\gamma d_2 (1-\beta \tan \alpha)$  &  $\gamma \frac{d_2}{c}(\beta -\tan \alpha)$ \\
   &  &  &  &  \\  
Q & 0 & 0 & 0 & 0 \\
   &  &  &  &  \\
R &  -$d_1$ & $\frac{d_1}{c} \tan \alpha$  &
  -$\gamma d_1 (1-\beta \tan \alpha)$  &  -$\gamma \frac{d_1}{c}(\beta -\tan \alpha)$ \\
   &  &  &  &  \\    
\hline
\end{tabular}\caption{ Space-time points on the object $O_D$, illuminated by a short light pulse
from the lamp L1 (Fig.2), as observed in the frames S',S.    }      
\end{center}
\end{table}
\begin{table}
\begin{center}
\begin{tabular}{|c|c c c c|} \hline
Point on object &  x' &  t' &  x  & t  \\ \hline
   &  &  &  &  \\
P &  $d_2$ & 0 & $\gamma d_2$  &  $\frac{\gamma v d_2}{c^2}$ \\
   &  &  &  &  \\
Q &0  & 0 & 0 & 0 \\
   &  &  &  &  \\
R &  -$d_1$ & 0 & -$\gamma d_1$  &  -$\frac{\gamma v d_1}{c^2}$ \\
   &  &  &  &  \\ 
\hline
\end{tabular}
\caption{Space-time points on the object $O_D$, illuminated by a short light pulse
from the lamp L2 (Fig.2), as observed in the frames S',S.   }      
\end{center}
\end{table}
\par As discussed in detail below, the time resolved image is, just as in the case
 of a stationary object, a vertical line moving parallel to O$x_I$. The meaning of
 `coarse' time resolution will also be quantified.
 \par Including both the effects of the LT and of the light propagation delays, $\Delta t_1$,
 $\Delta t_2$ are given, in the case that $O_D$ is moving transversely, by the equations :
 \begin{eqnarray}
 \Delta t_1^{L1} & = & -\frac{\gamma d_1}{c}(\tan \alpha-\beta)+\frac{d_1}{c} \frac
 {\{2 \tan \alpha -r_1[\gamma^2(1-\beta \tan \alpha)^2 +\tan^2\alpha] \}}
 {1+\sqrt{\gamma^2 (1-\beta \tan \alpha)^2 r_1^2 +(1-r_1 \tan \alpha)^2}}  \\
 \Delta t_2^{L1} & = & -\frac{\gamma d_2}{c}(\tan \alpha-\beta)+\frac{d_2}{c} \frac
 {\{2 \tan \alpha + r_2[\gamma^2(1-\beta \tan \alpha)^2 +\tan^2\alpha] \}}
 {1+\sqrt{\gamma^2 (1-\beta \tan \alpha)^2 r_2^2 +(1+r_2 \tan \alpha)^2}}    \\
\Delta t_1^{L2} & = & \frac{d_1 \gamma \beta }{c}+ \frac
 {d_1[2 \tan \alpha -r_1(\gamma^2+\tan^2 \alpha)]}
 {1+\sqrt{\gamma^2 r_1^2 +(1-r_1 \tan \alpha)^2}}    \\
\Delta t_2^{L2} & = & \frac{d_2 \gamma \beta }{c}+ \frac
 {d_2[2 \tan \alpha +r_2(\gamma^2+\tan^2 \alpha)]}
 {1+\sqrt{\gamma^2 r_2^2 +(1+r_2 \tan \alpha)^2}}    
 \end{eqnarray}
 where
 \[ r_1 = d_1/l,~~~~r_2 = d_2/l  \]
 On setting $\beta=0, \gamma =1$ in Eqns(3.5),(3.6) Eqns(2.2),(2.3) are recovered.
 In Eqns(3.3)-(3.6) the first terms on the right hand sides are the time shifts due
 to the LT, while the second terms account for light propagation delays between the
 object and the camera. The $\gamma, \beta$ dependent terms in the latter are also
 a consequence of the LT, that changes the apparent size of the object (see
 the fourth column of Tables 1 and 2) and so effects also the light propagation 
 delays.
\begin{figure}[htbp]
\begin{center}\hspace*{-0.5cm}\mbox{
\epsfysize10.0cm\epsffile{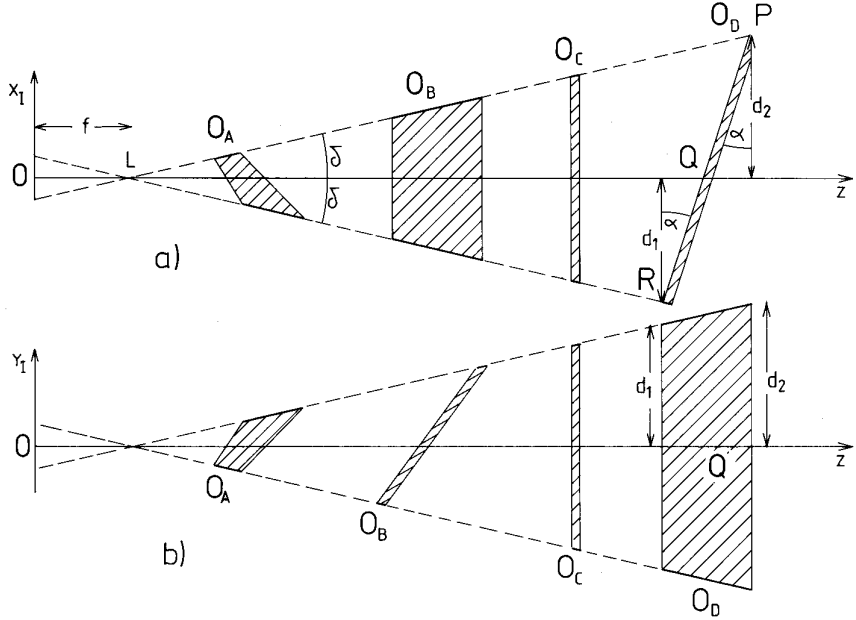}}
\caption{ a) $x_I$-$z$ projections, b) $y_I$-$z$ projections of the four planar
objects and the camera.} 
\label{fig-fig4}
\end{center}
\end{figure}  
\begin{figure}[htbp]
\begin{center}\hspace*{-0.5cm}\mbox{
\epsfysize20.0cm\epsffile{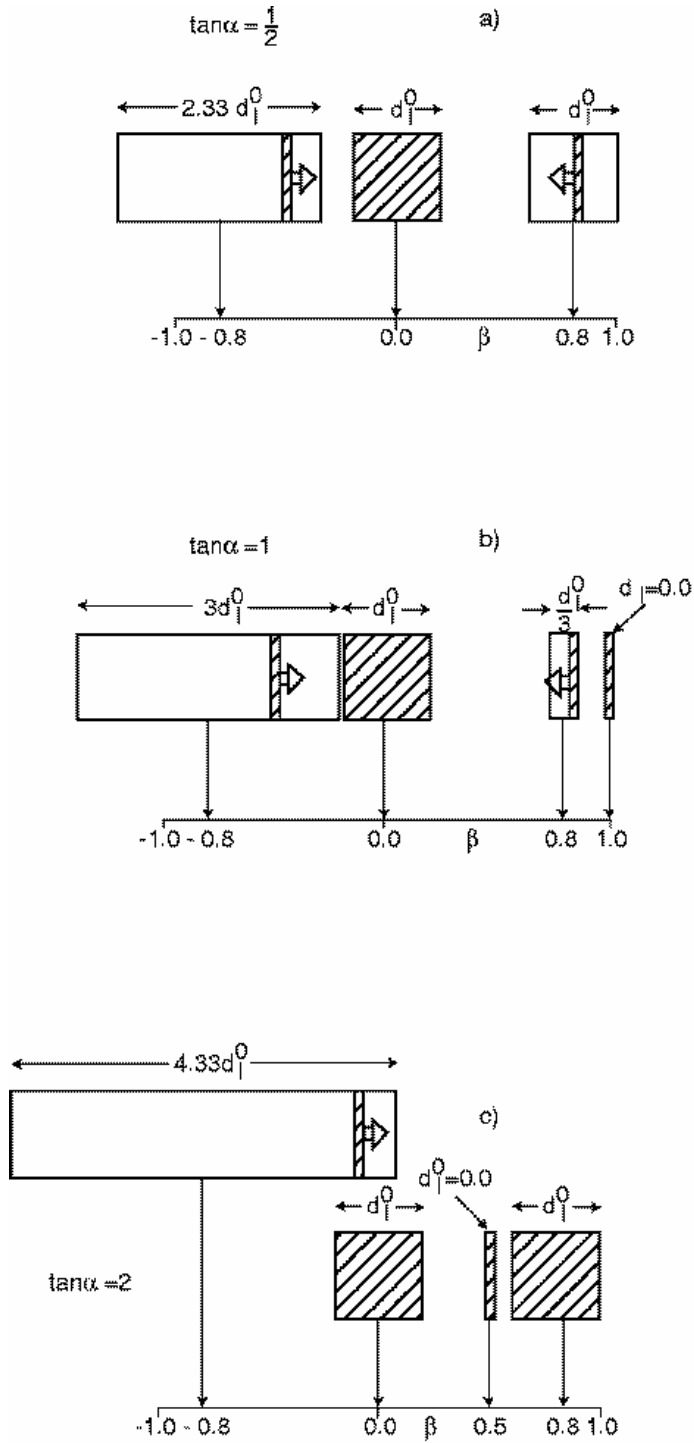}}
\caption{Images of $O_D$ observed when the lamp L1 is flashed while L1 and $O_D$
are in uniform motion relative to the camera with velocity $\beta c$ parallel to
 $O_{x_I}$. a),b),c) correspond to $\tan \alpha = 1/2, 1, 2$ respectively. Comments as for
Fig.3.} 
\label{fig-fig5}
\end{center}
 \end{figure}
\begin{figure}[htbp]
\begin{center}\hspace*{-0.5cm}\mbox{
\epsfysize22.0cm\epsffile{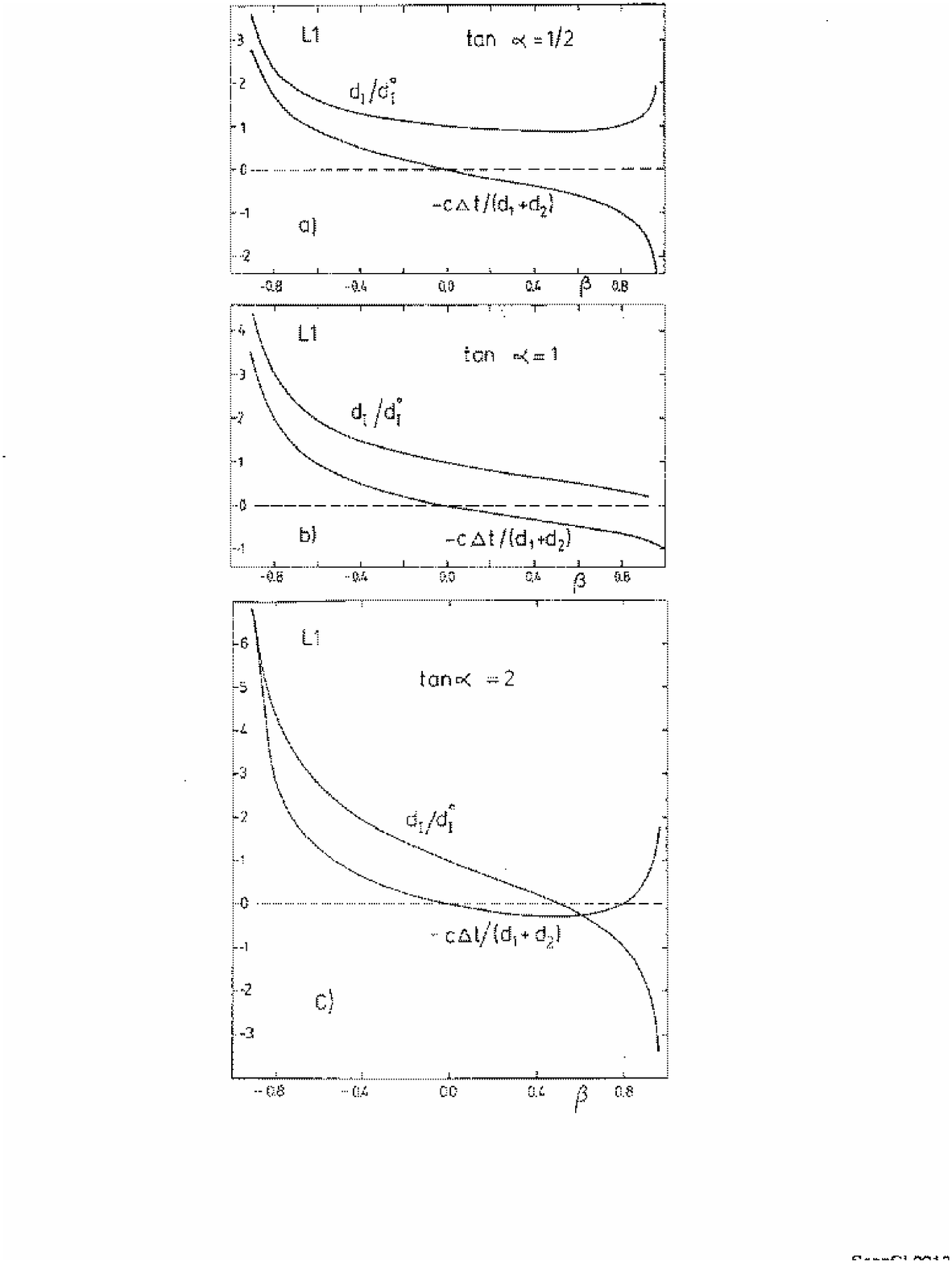}}
\caption{ $d_I/d_I^0$ and $-c \Delta t /(d_1+d_2)$ as a function of $\beta$ 
for the conditions of Fig.5. $d_I$ is the width parallel to $O_{x_I}$ of the
 rectangle in Fig.5, i.e. the width of the image when observed with coarse time
 resolution. $\Delta t$ is the total duration of the moving image. a),b),c) are for 
  $\tan \alpha = 1/2, 1, 2$ respectively.}
\label{fig-fig6}
\end{center}
 \end{figure}
 \par The images observed in the camera are now discussed in two extreme limits. In the first
 a similar approximation is made to that used in deriving Eqns(2.5),(2.6) from (2.2),(2.3).
 All terms containing $r_1$, $r_2$ in Eqns (3.3)-(3.6) are neglected. This is a good
 approximation for not too large $\gamma$ (say $\beta \le 0.9, \gamma \le 2$) and
 $ \tan \alpha,~ d_1,~d_2 \ll l$. The second is the ultrarelativistic limit $\gamma \gg 1$.
 In the small $\gamma,~ \tan \alpha$ limit Eqns (3.3)-(3.6) simplify to :
 \begin{eqnarray}
 \Delta t_i^{L1} & = & -\frac{d_i}{c}[(\gamma-1) \tan \alpha -\gamma \beta ]~~~~~i=1,2 \\
 \Delta t_i^{L2} & = & \frac{d_i}{c}[ \tan \alpha +\gamma \beta ]~~~~~i=1,2  
 \end{eqnarray}
 It is convenient to introduce the quantities $d_I$ for the width of the image
 observed with coarse time resolution and $\Delta t = \Delta t_1 + \Delta t_2$
 the full time over which the moving image exists. The quantity
  $\beta_I = d_I/c \Delta t$ is the average velocity, in units of c, of the
  moving line image. From Tables 1 and 2 and Eqns(3.7), (3.8) these quantities
  are:
  \begin{eqnarray}
  d_I^{L1} & = & \gamma (1-\beta \tan \alpha) d_I^0  \\
  d_I^{L2} & = & \gamma d_I^0 \\ 
  \Delta t^{L1} & = & -\frac{(d_1+d_2)}{c}[(\gamma - 1) \tan \alpha - \beta \gamma ] \\
  \Delta t^{L2} & = & \frac{(d_1+d_2)}{c}[ \tan \alpha + \beta \gamma ] 
  \end{eqnarray}
  where
  \begin{equation}
  d_I^0 = f \frac{(d_1+d_2)}{l} 
  \end{equation}
  The quantity $d_I^0$, introduced in the previous section, is the side of the square
  swept out by each moving image when $\beta = 0$ (see Fig 3). Typical images for L1 for
   different values of $\beta$ and for $\tan \alpha = 1/2,~1,~2$ are shown in Figs 5a,b,c 
   respectively. The corresponding plots of the quantities $d_I/d_I^0$ and $-c\Delta t
   /(d_1+d_2)$ as a function of $\beta$ are shown in Figs 6a,b,c. The convention in Fig 5 is
   the same as in Fig 3, the images are viewed from the object side and the dashed rectangles
   or squares indicate the full areas swept out by the moving line images. It is evident
   from Figs 5 and 6 that the images have a complicated structure as a function of
   $\beta$, but qualitatively they do not differ from those of a similar
    object at rest. For $\beta > 0$ ($< 0$) the line image moves to the left
     (right). For $\tan \alpha < 1$ $d_I^{L1}$ has a minimum value when 
     $\beta = \tan \alpha$ :
 \begin{equation}
 d_I^{L1,MIN} = d_I^0\sqrt{1-\tan^2 \alpha}
  \end{equation}
  while for $\tan \alpha > 1$, $\Delta t$ has a maximum when
  $\beta = 1 / \tan \alpha$ :
 \begin{equation}
 \Delta t^{L1,MAX} = \frac{(d_1+d_2)}{c}\left[ \frac{\sqrt{\tan ^2 \alpha-1}}
 {\tan \alpha}-\tan \alpha \right] 
  \end{equation}          
  For $\tan \alpha >1$, $d_I$ vanishes for some positive value of $\beta$
  ( Figs. 5c,6c), and $\Delta t$ has a second zero 
  for positive $\beta$ (by construction, it of course vanishes for 
  $\beta = 0$ and any $\alpha$).
  When $\tan \alpha = 2$, $\Delta t$ vanishes for $\beta = 0.8$. The instantaneous
  image is then identical to that obtained for $\beta = 0$. It is easily shown, using
  Eqns.(3.9),(3.11), that the same instantaneous image is obtained when
   $\Delta t = 0$ for any $\tan \alpha >1$.
   Note that, because of
  the approximations made above, Eqns.(3.9),(3.11) are not valid for 
  $\beta \simeq 1$ (see the discussion of the ultra-relativistic limit
  where $|\beta| \simeq 1$ below).
  \par The structure of the images is simpler for the case of the lamp L2
  (Figs.7 and 8 ). $d_I^{L2}$ has the constant value $\gamma d_I^0$ independant
   of $\alpha$. Since this is an example of Space Dilatation the only 
   relevant parameter is the total extension of the object along the Ox'
   axis. For $\gamma \beta > -\tan \alpha$  ($\gamma \beta < -\tan \alpha$)
   the line image moves to the left (right), see Fig 7. At
    $\gamma \beta = -\tan \alpha$ ( Figs. 7,8) then $\Delta t= 0$ and an 
    instantaneous rectangular image is seen. The minimum value of $d_I$ of
    $d_I^0$ occurs when $\beta = 0$. $\Delta t$ is an increasing monotonic
    function of $\beta$ for all values of $\beta$.
\begin{figure}[htbp]
\begin{center}\hspace*{-0.5cm}\mbox{
\epsfysize10.0cm\epsffile{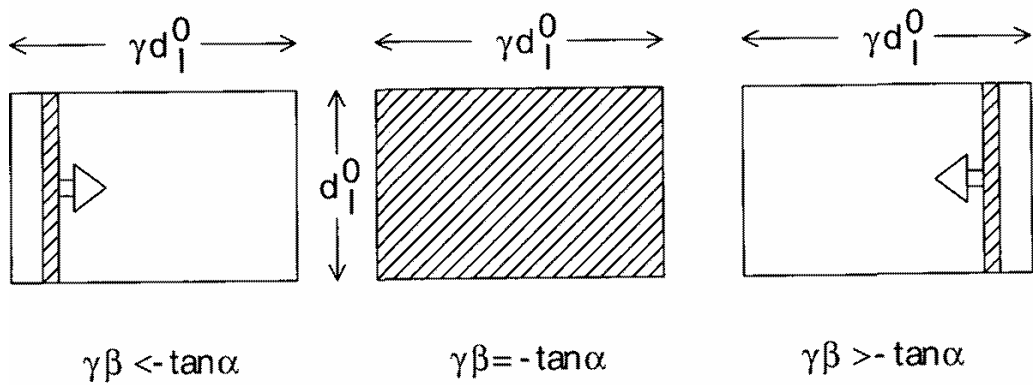}}
\caption{ Images of $O_D$ observed when lamp L2 is flashed and L2 and $O_D$
move uniformly relative to the camera with velocity $\beta c$ parallel to $O_{x_I}$. 
 Comments as for Fig.3.} 
\label{fig-fig7}
\end{center}
 \end{figure}
\begin{figure}[htbp]
\begin{center}\hspace*{-0.5cm}\mbox{
\epsfysize10.0cm\epsffile{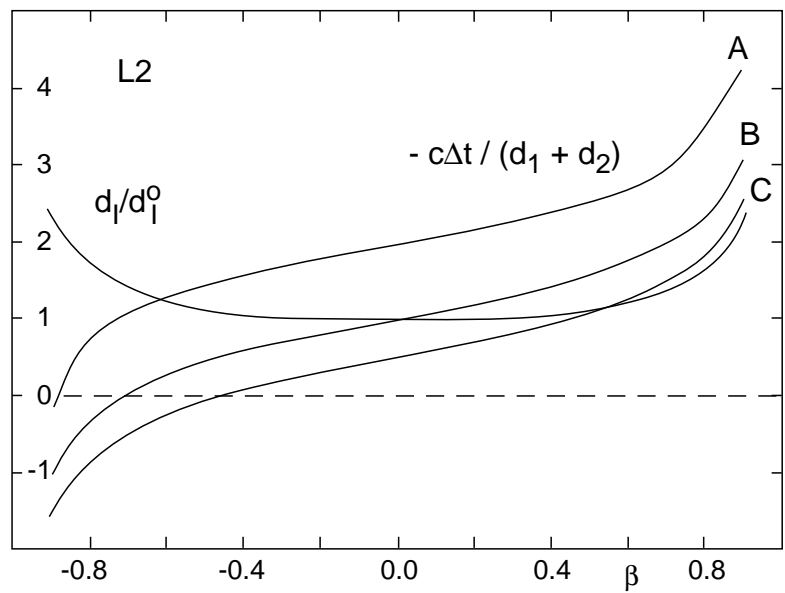}}
\caption{$d_I/d_I^0$ and $-c \Delta t/(d_1+d_2)$ (Curves A, B, C) as a function of $\beta$
 for the conditions of Fig.7 (lamp L2 flashed). Curves A,B,C for 
 $\tan \alpha = 1/2, 1, 2$ respectively.}
\label{fig-fig8}
\end{center}
 \end{figure} 
    \par The form of the images in the Ultra-Relativistic (UR) limit 
    $|\beta| \simeq 1$ will now be discussed. In this case only the $\gamma$ dependent terms in 
    Eqns.(3.3)-(3.6) are retained, leading to the expressions:
 \begin{eqnarray}
 \Delta t^{L1} & = & \frac{\gamma}{c}[(d_1+d_2)(\beta - \tan \alpha) + (d_1+d_2)(\beta \tan \alpha
 -1)] \\  
 \Delta t^{L2} & = & \gamma [ d_2(1+\beta)-d_1(1-\beta)]
 \end{eqnarray}
 while Eqns.(3.9),(3.10), being exact, remain valid. For lamp L1 different behaviour in the
 UR limit occurs for $\tan \alpha \neq \pm 1$ and  $\tan \alpha = \pm 1$. For
 $\tan \alpha \neq \pm 1$, Eqns.(3.9),(3.16) give
  $|d_I^{L1}|$, $|\Delta t^{L1}|  \rightarrow \infty$ as
 $|\beta| \rightarrow 1$. The velocity $\beta_I$ is however finite and independent,
  for $|\beta| =1$, of  $\tan \alpha$. In general, in the UR limit :
 \begin{equation}
 \beta_I^{L1} = \frac{d_I}{c \Delta t} = \frac{(1-\beta \tan \alpha) d_I^0}
 {[(d_1+d_2)(\beta - \tan \alpha)+(d_1-d_2)(\beta \tan \alpha -1)]}
 \end{equation}
 so
\begin{equation}
 \beta_I^{L1}(\beta = 1) = \frac{f}{2l}(1+\frac{d_1}{d_2})
 ~~~(\tan \alpha \neq \pm 1)
 \end{equation} 
 and
\begin{equation}
 \beta_I^{L1}(\beta = -1) = -\frac{f}{2l}(1+\frac{d_2}{d_1})
 ~~~(\tan \alpha \neq \pm 1) 
 \end{equation}  
 For $\tan \alpha = 1$, (3.7), (3.16) give 
 \begin{eqnarray}
 d_I^{L1}(\tan \alpha =1) & = & \sqrt{\frac{1-\beta}{1+\beta}} d_I^0  \\ 
 \Delta t^{L1}(\tan \alpha =1) & = & -\frac{2}{c}\sqrt{\frac{1-\beta}{1+\beta}} d_1
 \end{eqnarray}
 so that, with Eqn.(3.13):
\begin{equation}
 \beta_I^{L1}( \tan \alpha =1) = -\frac{f}{2l}(1+\frac{d_2}{d_1})
 \end{equation}  
 whereas for $\tan \alpha = -1$, one obtains:
\begin{eqnarray}
 d_I^{L1}(\tan \alpha = -1) & = & \sqrt{\frac{1+\beta}{1-\beta}} d_I^0  \\ 
 \Delta t^{L1}(\tan \alpha = -1) & = & \frac{2}{c}\sqrt{\frac{1+\beta}{1-\beta}} d_2
 \end{eqnarray}
 and hence
\begin{equation}
 \beta_I^{L1}( \tan \alpha = -1) = \frac{f}{2l}(1+\frac{d_1}{d_2})
 \end{equation}   
 Thus $d_I^{L1}$ and $\Delta t^{L1}$, for $\tan \alpha = 1$, vanish as $\beta \rightarrow 1$
 and are infinite as $\beta \rightarrow -1$. On the contrary, $d_I^{L1}$ and $\Delta t^{L1}$,
  for $\tan \alpha = -1$, are infinite as $\beta \rightarrow 1$
 and vanish as $\beta \rightarrow -1$. For $\tan \alpha = 1$ the limiting velocity is the same
 as for $\tan \alpha \neq 1$ and $\beta =  -1$, while for $\tan \alpha = -1$ it is the same as for
 $\tan \alpha \neq 1$ and $\beta = 1$.
 The UR limits for lamp L1 in all the cases discussed above are summarised
 in Table 3.
\begin{table}
\begin{center}
\begin{tabular}{|c|c c|c c|c c|} \hline
   & \multicolumn{2}{|c|}{$\tan \alpha \ne \pm 1$}
   & \multicolumn{2}{c|}{$\tan \alpha = 1$}
   & \multicolumn{2}{c|}{$\tan \alpha = -1$} \\ \cline{2-7}
   & \multicolumn{1}{|c|}{$\beta \rightarrow 1$}
   & \multicolumn{1}{c|}{$\beta \rightarrow -1$}
   & \multicolumn{1}{c|}{$\beta \rightarrow 1$}
   & \multicolumn{1}{c|}{$\beta \rightarrow -1$}
   & \multicolumn{1}{c|}{$\beta \rightarrow 1$}
   & \multicolumn{1}{c|}{$\beta \rightarrow -1$} \\ \cline{1-7}
   &  &  &  &  &  &  \\
 $|d_I^{L1}|$ & $\infty$ &  $\infty$ &  0
 & $\infty$ &  $\infty$ &  0 \\
   &  &  &  &  &  &  \\ 
 $|\Delta t^{L1}|$ & $\infty$ &  $\infty$ &  0
 & $\infty$ &  $\infty$ &  0 \\
   &  &  &  &  &  &  \\ 
 $\beta_I^{L1}$ & $\frac{f}{2l}(1+\frac{d_1}{d_2})$ &
 $-\frac{f}{2l}(1+\frac{d_2}{d_1})$ &
 $-\frac{f}{2l}(1+\frac{d_2}{d_1})$ &
 $-\frac{f}{2l}(1+\frac{d_2}{d_1})$ &
 $\frac{f}{2l}(1+\frac{d_1}{d_2})$ &
 $\frac{f}{2l}(1+\frac{d_1}{d_2})$  \\
   &  &  &  &  &  &  \\   
\hline
\end{tabular}
\caption[]{ The size $|d_I^{L1}|$, time duration $|\Delta t^{L1}|$ and velocity
 parallel to O$x_I$ $\beta_I^{L1}$, of images of the object $O_D$,
  illuminated by the lamp L1 for
 different values of $\tan \alpha$ in the UR limit.  }      
\end{center}
\end{table} 
\begin{table}
\begin{center}
\begin{tabular}{|c|c c|} \hline
   & \multicolumn{1}{c|}{$\beta \rightarrow 1$}
   & \multicolumn{1}{c|}{$\beta \rightarrow -1$} \\ \cline{1-3}
   &  &  \\
   $d_I^{L2}$ &  $\infty$ & $\infty$ \\
   &  &  \\   
   $\Delta t^{L2}$ &  $\infty$ & $-\infty$ \\
   &  &  \\    
 $\beta_I^{L2}$ & $\frac{f}{2l}(1+\frac{d_1}{d_2})$ &
 $-\frac{f}{2l}(1+\frac{d_2}{d_1})$ \\
   &  &  \\        
\hline
\end{tabular}
\caption[]{ The size $|d_I^{L2}|$, time duration $|\Delta t^{L2}|$ and velocity
(parallel to O$x_I$) $\beta_I^{L2}$, of images of the object $O_D$,
  illuminated by the lamp L2 in the UR limit.    }      
\end{center}
\end{table}
 \par As there is no $\tan \alpha$ dependence the UR limits for the lamp L2 are
 simpler. For $|\beta| \rightarrow 1$, $d_I^{L2}$  always diverges to 
 $+\infty$, whereas $\Delta t^{L2}$ diverges to $+\infty$ as $\beta \rightarrow 1$
 and to $-\infty$ as $\beta \rightarrow -1$. The same limiting velocities are found as
 $\beta \rightarrow \pm 1$ as 
 for lamp L1 when $\tan \alpha = \mp 1$. These results are summarised in Table 4.
 It may be remarked that for $\tan \alpha \neq \pm 1$ lamps L1 and L2 give similar, 
 infinitely wide, images in the UR limit.
 \par In the above discussion optical aberration has been neglected and it is assumed that
 the camera is still sensitive for large values of $\beta$ where the observed photons 
 have a large red-shift. Denoting by $\theta$, $\theta$' the angles in the S, S' frames 
 of the diffusely reflected photons relative to the x, x' axes, the LT of the photon 
 momentum leads to the relation :
\begin{equation}
\tan \theta = \frac{ \sin \theta'}{\gamma (\cos \theta'+\beta)} 
\end{equation}
In order to enter the camera the photons must have $\theta \simeq \frac{\pi}{2}$,
 implying from Eqn.(3.27) that $\cos \theta' \simeq -\beta$. If $\theta' > \pi-\alpha$,
 then (see Fig2.) the photons cannot be diffusely reflected from the surface of the object.
 This leads, for any value of $\alpha$, to a maximum value of $\beta$ for diffuse reflection.
 For $\tan \alpha = 1/2,~1,~2$ as in Figs.5,6  $\beta_{MAX}=0.88,~0.707,~0.440$.
 If the object is translucent, or equipped with a lamp similarly situated to L1, but
 illuminating the back surface of the object (i.e. placed at
 the position the mirror image of 
 L1 in the object) then the above limitation is avoided. For $\beta > \beta_{MAX}$ 
 the photons entering the camera originate entirely from reflection on the surface 
 remote from the camera. This is just the optical aberration effect referred to in the
 Introduction that leads to the apparent rotation of of  rapidly moving objects
 ~\cite{x2,x3,x4}. No such restriction applies for negative values of $\beta$ and in
 this case the photons are always reflected from the front surface of the object.
 \par Since (compare Figs.3,5,7) the images of stationary and moving TLO are not 
 distiguishable, and the moving image is defined by just three independent 
 quantities (say $\Delta t_1$, $\Delta t_2$ and $d_I$) whereas four
  (for example $d_1$, $\alpha$, $l$ and $v$ ) are needed to completely specify
   the moving planar object, image plane measurements using only one lamp
   do not enable reconstruction of the object. It suffices however to perform
   measurements, under identical conditions, using separately L1 and L2 to derive the 
   true dimensions, orientation, position and velocity of the object\footnote{
   In general equivalent information is provided by any two lamps situated in the 
   plane defined by the optic axis of the camera and the normal to the surface of the 
   object at different known angular positions with respect to
   the normal.}. From Eqns.(3.9,3.10):
\begin{equation}
r_d \equiv \frac{d_I^{L1}}{d_I^{L2}} = 1 -\beta \tan \alpha  
\end{equation}         
and from Eqns.(3.11),(3.12)
\begin{equation}
r_t \equiv \frac{\Delta t^{L1}}{ \Delta t^{L2}} = \frac{\beta \gamma -(\gamma-1) \tan \alpha}
{\tan \alpha + \beta \gamma}
\end{equation}
Eliminating $\tan \alpha$ between Eqns.(3.28) and (3.29) and solving for $\gamma$
gives:
\begin{equation}  
\gamma = \sqrt{r_3^2+r_4}-r_3
\end{equation}
where
\[ r_3 \equiv \frac{(1-r_d)(1-r_t)}{2(r_d-r_t)},~~~~ r_4 \equiv \frac{(1-r_t)}{r_d-r_t}\]
Re-writing Eqn.(3.28), $\tan \alpha$ is then given by:
\begin{equation} 
\tan \alpha = \frac{\gamma}{\sqrt{\gamma^2-1}}(1-r_d)
\end{equation}
Since $\tan \alpha$, $\beta$, $\gamma$ are now known, $d_1$ and $d_2$ may be determined from 
Eqns.(3.7) or (3.8). Finally, $l$ is given by the relation: 
\begin{equation}
l = \frac{2d_1 d_2 \tan \alpha}{d_2-d_1}
\end{equation}
Thus by combining measurements made under similar conditions, using lamps L1 and L2 a
complete specification of the moving object is obtained using only the IPC information.
\SECTION{\bf{Space Time Measurements of Equivalent Moving Clocks}}
In this Section space time measurements of an array of synchronised clocks situated in the 
inertial frame S' will be considered. These clocks may be synchronised by any convenient
procedure \footnote{If an observer in S' knows the distance $D$ to any of the clocks
 then the clock is synchronised relative to a local clock at the same position as the 
 observer, when it is observed to lag behind the latter by the time $D/c$ when viewed across
 free space}
 (see for example Ref.[1]). For an observer in S' all such clocks are `equivalent' in the sense
 that each of them records, independently of its position, the proper time $\tau'$ of the frame
 S'. For convenience, the array of clocks is assumed to be placed on the wagons of a
 train which is at rest in S', as shown in Fig.9a. The clocks are labelled $C_m~, m=...-2,-1,
 0,1,2,...$ and are situated (with the exception of the `magic clocks' $C_M,
  \overline{C}_M$,
 see below) at fixed distances $L$ from each other, along the Ox' axis, which is parallel 
 to the train. Any observer in S' will, after making the necessary corrections for Light 
 Propagation Time Delays (LPTD), note that each Equivalent Clock (EC) indicates the same time,
 as shown in Fig.9a. It is now asked how the array of EC will appear to an observer at a fixed
 position in the frame S when the train is moving with velocity $\beta c$ parallel to the
 direction Ox in S (Fig.9b). It is assumed that the EC $C_0$ is placed at $x'=0$ and 
 that it is synchronised with the Standard Clock $C_S$, placed at $x=0$ in S, when
 $t=t'=0$. The EC $C_m$ and $C_S$ record exactly equal time intervals when they are 
 situated in the same inertial frame.
\begin{figure}[htbp]
\begin{center}\hspace*{-0.5cm}\mbox{
\epsfysize10.0cm\epsffile{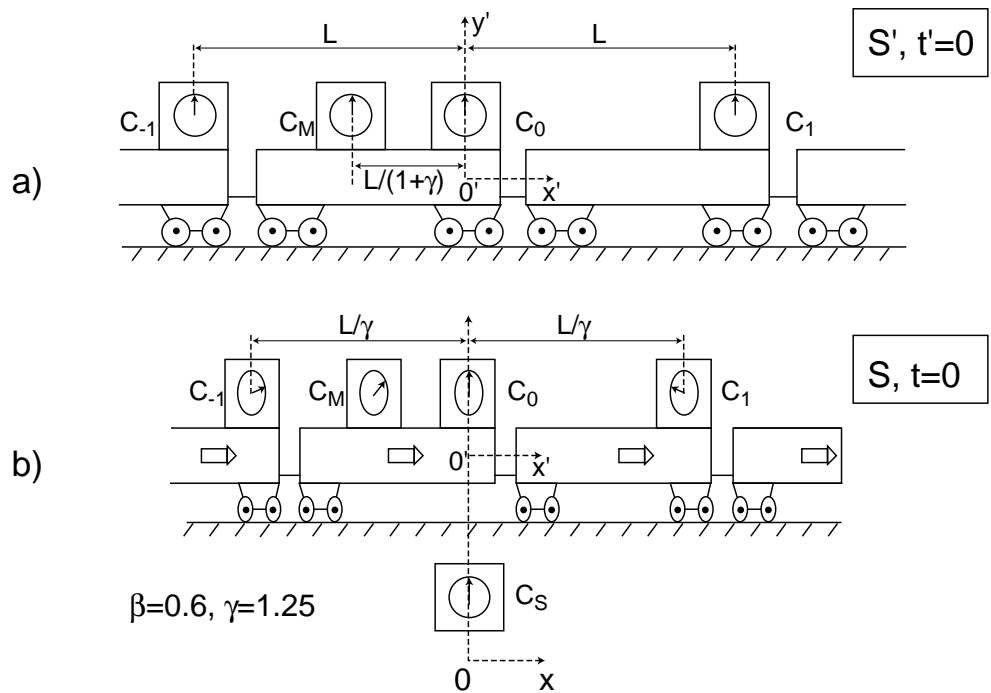}}
\caption{a) Positions and times of equivalent clocks on the wagons
of a train as seen by observers in the rest frame S' of the train
(without the effects of LPTD).
b) The positions and times of the same clocks as seen by an observer in S
(without the effects of LPTD). In S the train is moving to the right with
velocity $\beta c$.}
\label{fig-fig9}
\end{center}
 \end{figure}    
\begin{figure}[htbp]
\begin{center}\hspace*{-4.5cm}\mbox{
\epsfysize12.0cm\epsffile{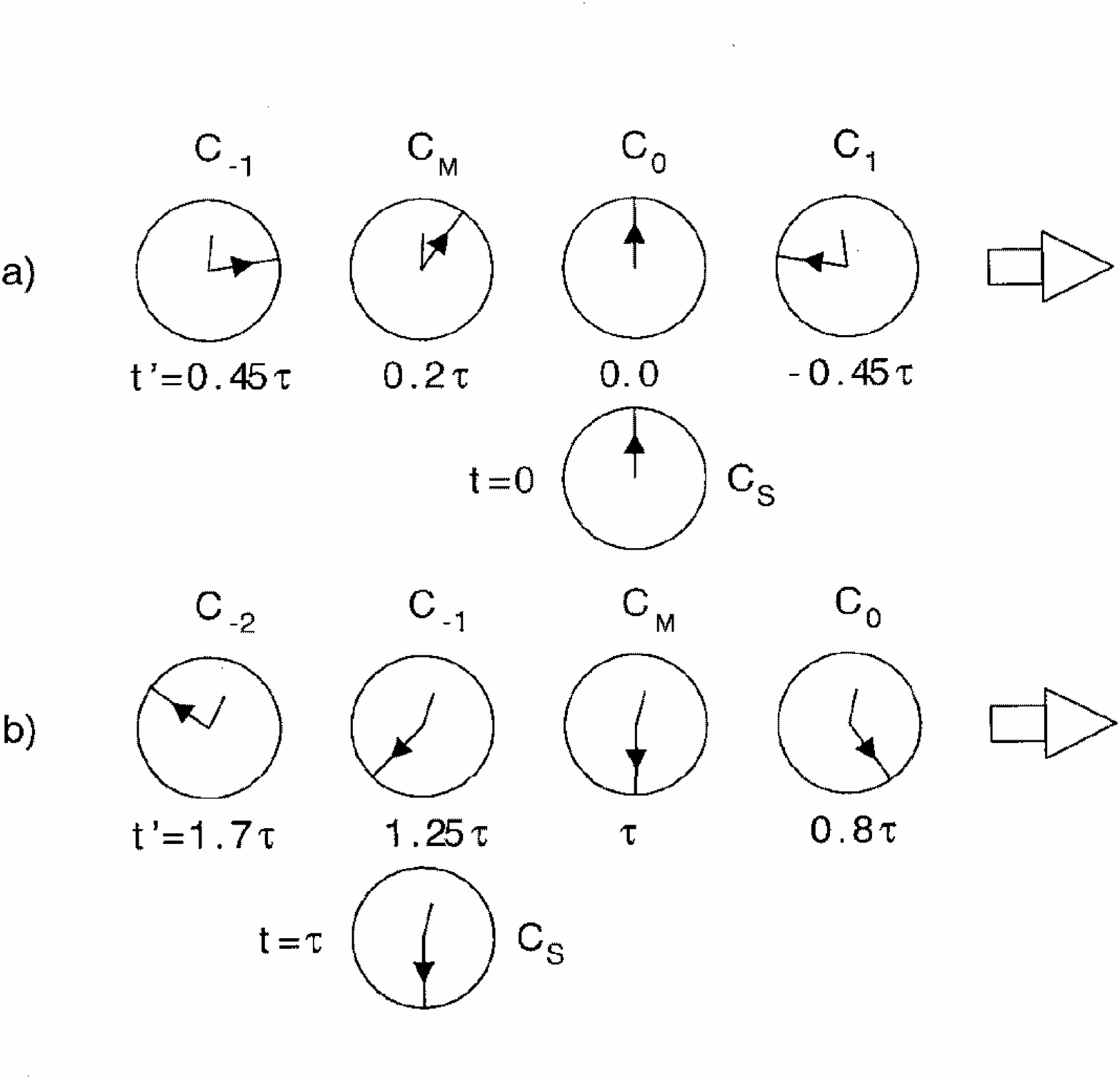}}
\caption{ Equivalent clocks on the train as seen by an observer in S.
a) at $t = 0$, b) at $t = \tau$ (without the effects of LPTD).}
\label{fig-fig10}
\end{center}
 \end{figure} 
 \par The appearence of the moving EC to an observer in S (after correction for LPTD; their actual
 appearence, including this effect, is considered later) at $t=0$ is shown in Fig.9b, and in
 more detail in Fig.10 for both $t=0$ and $t=\tau$. The period $\tau$ is the time between
  the passage of successive EC past $C_S$. The big hand of $C_S$ in Fig.10 rotates through
  $180^{\circ}$ during the time $\tau$. Explicit expressions for the apparent times are
  presented in Table 5. In Fig.9b,10 the times indicated by
  the clocks are shown 
  for $\beta= 0.6$. These apparent times are readily calculated using the LT equations
  (3.1),(3.2). Consider the time indicated by $C_1$ at $t=0$. The space-time points are:
  \[~~~S'~:~(L,t')~~;~~~S~:~(x,0) \] 
  Hence, Eqns.(3.1),(3.2) give:
  \begin{eqnarray}
  x & = & \gamma (L + vt')   \\
  0 & = & \gamma (t'+\frac{\beta L}{c})
  \end{eqnarray}
  which have the solution [ $C_1(t=0)$ ]:
\begin{eqnarray}
  t' & = & -\frac{\beta L}{c}   \\
  x & = & \frac{L}{\gamma}
  \end{eqnarray} 
\begin{table}
\begin{center}
\begin{tabular}{|c|c c c c c c|} \hline
 $C_S$  & \multicolumn{1}{c|}{ $C_{-2}$}
   & \multicolumn{1}{c|}{ $C_{-1}$}
   & \multicolumn{1}{c|}{ $C_{M}$} 
   & \multicolumn{1}{c|}{ $C_{0}$}
   & \multicolumn{1}{c|}{ $C_{1}$}
   & \multicolumn{1}{c|}{ $C_{2}$} \\ \cline{1-7}
   &  &  &  &  &  &  \\   
 0 & $2 \frac{(\gamma^2-1)}{\gamma} \tau$ 
   & $ \frac{(\gamma^2-1)}{\gamma} \tau$
   & $ \frac{(\gamma -1)}{\gamma} \tau$ & 0 
   & $ -\frac{(\gamma^2-1)}{\gamma} \tau$
   & $ -2 \frac{(\gamma^2-1)}{\gamma} \tau$ \\
   &  &  &  &  &  &  \\   
$\tau$ & $ \frac{(2 \gamma^2-1)}{\gamma} \tau$ 
   & $ \gamma \tau$
   & $\tau$ 
   & $\frac{\tau}{\gamma}$
   & $ -\frac{(\gamma^2-2)}{\gamma} \tau$
   & $ -\frac{(2 \gamma^2-3)}{\gamma} \tau$ \\
   &  &  &  &  &  &  \\             
\hline
\end{tabular}
\caption[]{Apparent times of Equivalent Clocks on the moving train in Fig.9, 
at times $t=0$ and $t=\tau$ of the stationary standard clock $C_S$. Effects of
the LT only.  }      
\end{center}
\end{table}
\begin{table}
\begin{center}
\begin{tabular}{|c|c c c c c c|} \hline
 $C_S$  & \multicolumn{1}{c|}{ $C_{-2}$}
   & \multicolumn{1}{c|}{ $C_{-1}$}
   & \multicolumn{1}{c|}{ $\overline{C}_{M}$} 
   & \multicolumn{1}{c|}{ $C_{0}$}
   & \multicolumn{1}{c|}{ $C_{1}$}
   & \multicolumn{1}{c|}{ $C_{2}$} \\ \cline{1-7}
   &  &  &  &  &  &  \\   
 0 & $-2 \gamma \beta \tau$
   & $ - \gamma \beta \tau$  
   & $\left[ \sqrt{(1-\beta)/(1+\beta)}-1\right] \tau$ & 0
   & $ - \gamma \beta \tau$ 
   & $-2 \gamma \beta \tau$ \\
   &  &  &  &  &  &  \\   
 $\tau$ & $\gamma (1-\beta) \tau$ 
   & $\gamma \tau$ & $\tau$ 
   & $\gamma (1-\beta) \tau$ 
   & $\gamma (1- 2 \beta) \tau$
   & $\gamma (1- 3 \beta) \tau$ \\
   &  &  &  &  &  &  \\           
\hline
\end{tabular}
\caption[]{ Definitions as for Table 5, except that the effects of LPTD for an
observer close to the train are also included.   }      
\end{center}
\end{table}   
  As shown in Fig 9b, the wagons of the train are apparently shorter due to the LFC
  effect (Eqn.(4.4)) and also {\it the wagons at the front end of the train
  are seen at an earlier time than those at the rear end}. Thus a $t=0$ snapshot
   in S corresponds, not to a fixed $t'$ in S' but one which depends on $x'$: 
   $t'=-\beta x'/c$. This is a consequence of the relativity of simultaneity of space-time
   events in S and S', as first pointed out by Einstein in Ref.[1]. Here it appears in a 
   particularly graphic and striking form. The part of the space-time domain in S' that may
   be observed from S is considered in detail below. Consider now the time indicated by
   $C_{-1}$ at $t=\tau$, i.e. when $C_{-1}$ is at the origin of S. The space-time points
   are:   
  \[~~~S'~:~(-L,t')~~;~~~S~:~(0,\tau) \] 
  Hence, Eqns.(3.1),(3.2) give:
  \begin{eqnarray}
  0 & = & \gamma (-L + vt')   \\
  \tau & = & \gamma (t'-\frac{\beta L}{c})
  \end{eqnarray}
  with the solutions [ $C_{-1}(t=\tau)$ ]:
\begin{eqnarray}
  t' & = & \frac{L}{v}   \\
  \tau & = & \frac{L}{\gamma v} = \frac{t'}{\gamma}
  \end{eqnarray}    
 so that
 \begin{equation}
 t'= \gamma \tau
 \end{equation}
  The EC at the origin of S at $t=\tau$ indicates a later time than $C_S$ i.e.
 it is apparently running {\it faster} than $C_S$. This is an example of Time Contraction (TC).
 As shown below {\it TC is exhibited by the EC at any fixed position in S}.
 In fact, if the observer in S can see the EC only when they are near to $C_S$ he (or she) will
 inevitably conclude that the clocks on the train run fast, not slow as in the classical TD
 effect (see below). Suppose that the observer is sitting in a waiting room with the clock
 $C_S$ and notices the time on the train (the same as $C_S$) by looking at $C_0$ as it passes 
 the waiting room window. If he (or she) then compares $C_{-1}$ as it passes the window with $C_S$
 it will be seen to be running fast relative to the latter. In order to see the TD effect
 the observer would (as will now be shown), have to note the time shown by, for example
 $C_0$, at time $t=\tau$ as recorded by $C_S$. Indeed, to do this he would first have to
 correct his observation for the LPTD between himself and $C_0$. At time $t=\tau$ at $C_S$
 he would actually see $C_0$ as it appeared at an earlier time to a nearby observer in S.
 Using Eqn.(4.8),Eqn.(4.3) may be written as [$C_1(t=0)$]:
 \begin{equation}
 t' = -\beta^2 \gamma \tau = -\frac{(\gamma^2-1)\tau}{\gamma}
 \end{equation}
 This is the formula for the apparent time reported in Table 5.
 Now consider $C_0$ at time $t=\tau$. The space-time points are:
 \[~~~S'~:~(0,t')~~;~~~S~:~(x,\tau) \] 
  Hence, Eqns.(3.1),(3.2) give:
  \begin{eqnarray}
  x & = & \gamma v t'   \\
  \tau & = & \gamma t'
  \end{eqnarray}
  with the solutions [ $C_0(t=\tau)$ ]:
\begin{eqnarray}
  t' & = & \tau/ \gamma \\
  x & = & v \tau  = L/\gamma
  \end{eqnarray}
 So {\it the EC $C_0$ at time $t=\tau$ indicates an earlier time, and so is 
 apparently running slower than $C_S$}.
 This is the classical Time Dilatation (TD) effect. It applies to observations
 of all {\it local clocks in S'},(i.e. those situated at a fixed value of $x'$) as well as any 
 other EC that has the same value of $x'$.
 \par As a last example consider the `Magic Clock' $C_M$ shown in Fig 9a at time $t=\tau$.
 With the space-time points:
 \[~~~S'~:~(-L/(1+\gamma),t')~~;~~~S~:~(x,\tau) \] 
  Eqns.(3.1),(3.2) give:
  \begin{eqnarray}
  x & = & \gamma [-L/(1+\gamma)+v t']   \\
  \tau & = & \gamma [ t'-\frac{\beta}{c}L/(1+\gamma)]
  \end{eqnarray}
  with the solutions [ $C_M(t=\tau)$ ]:
\begin{table}
\begin{center}
\begin{tabular}{|c|c c c c c c|} \hline
 $C_S$  & \multicolumn{1}{c|}{ $C_{-2}$}
   & \multicolumn{1}{c|}{ $C_{-1}$}
   & \multicolumn{1}{c|}{ $C_{M}$} 
   & \multicolumn{1}{c|}{ $C_{0}$}
   & \multicolumn{1}{c|}{ $C_{1}$}
   & \multicolumn{1}{c|}{ $C_{2}$} \\ \cline{1-7}
   &  &  &  &  &  &  \\   
 0 & $2 \frac{(\gamma^2-1)}{\gamma} \tau$ 
   & $ \frac{(\gamma^2-1)}{\gamma} \tau$
   & $ \frac{(\gamma -1)}{\gamma} \tau$ & 0 
   & $ -\frac{(\gamma^2-1)}{\gamma} \tau$
   & $ -2 \frac{(\gamma^2-1)}{\gamma} \tau$ \\
   &  &  &  &  &  &  \\   
$\tau$ & $ \frac{(2 \gamma^2-1)}{\gamma} \tau$ 
   & $ \gamma \tau$
   & $\tau$ 
   & $\frac{\tau}{\gamma}$
   & $ -\frac{(\gamma^2-2)}{\gamma} \tau$
   & $ -\frac{(2 \gamma^2-3)}{\gamma} \tau$ \\
   &  &  &  &  &  &  \\             
\hline
\end{tabular}
\caption[]{Apparent times of Equivalent Clocks on the moving train in Fig.9, 
at times $t=0$ and $t=\tau$ of the stationary standard clock $C_S$. Effects of
the LT only.  }      
\end{center}
\end{table}
\begin{eqnarray}
  t' & = & \tau  \\
  x & = & \gamma v \tau / (1+\gamma) 
  \end{eqnarray}
  where the relation $L=\gamma v \tau$ from Eqn.(4.14) has been used.
  {\it Thus $C_M$ indicates the same time as $C_S$ at $t=\tau$}. 
  Similar `Magic Clocks' can be defined that show the same time as $C_S$ at any chosen time 
  $t$ in S. Such a clock is, in general, situated at $x' = -ct(\gamma-1)/\beta \gamma$.
  All of the other apparent times presented in Table 5 and shown in Figs. 9b, 10 are calculated
  in a similar way to the above examples by choosing appropriate values of $x'$ and t.
  \par It is straightforward to derive a general formula for the apparent time of any EC, $C_m$
  after the passage of an arbitary number $j$ of wagons past the clock $C_S$. Still
  neglecting LPTD, the result is :
  \begin{equation}
  t'_{m,j} = -\frac{[m \gamma ^2-(m+j)] \tau}{\gamma}
  \end{equation}
  A consequence of (4.19) is :
  \begin{equation}
  t'_{m,j+1} - t'_{m,j} = \tau / \gamma
 \end{equation}
 This is the general TD result for any local ($x' =$ constant) EC $C_m$.
 Eqn.(4.19) may, alternatively, be written in terms of ($n$, $j$) where the index
 $n$ labels the position of an EC in S rather than in S'. So the clocks at
 $x = L/\gamma,~2L/\gamma,~...$ (see Fig. 9b) have $n=1,2,...$,those with 
  $x = -L/\gamma,~-2L/\gamma,~...$ have  $n=-1,-2,...$.
  Using the general relation :
  \begin{equation}
  n=m+j
  \end{equation}
  Eqn.(4.19) may be written as :
 \begin{equation}
  t'_{n,j} = -\frac{[(n-j)\gamma ^2-n] \tau}{\gamma}
  \end{equation}
  so that
  \begin{equation}
  t'_{n,j+1} - t'_{n,j} = \gamma \tau 
 \end{equation}    
 This is the general TC effect for an EC at fixed n ($x=$ constant).
 Eqn.(4.22) may also be used to calculate the apparent time delay between the EC
 on successive wagons at a fixed time in S :
 \begin{equation}
  t'_{n+1,j} - t'_{n,j} = -\gamma \beta^2 \tau 
 \end{equation} 
 \par The effects of LPTD on the apparent times indicated by the clocks on 
 the moving train will now be taken into account. Only propagation times 
 parallel to the train are considered and it is assumed that the clocks are
 orientated in such away that they can be seen by an observer placed beside
 $C_S$. Consider the clock $C_n$ at the time $\Delta t$ before the passage
 of the $j$th wagon past $C_S$ (Fig.11). If the clock m is at the position
 $x_m$ in S at this time then the inverse of the LT Eqn.(3.1) gives:
 \begin{equation}
 mL = \gamma [x_m-v(j \tau-\Delta t)]
 \end{equation} 
 The corresponding time shown by $C_m$, $t'^D_{m,j}$ is, using the inverse of
  the LT Eqn.(3.2):
  \begin{equation}
 t'^D_{m,j}  = \gamma [ j \tau - \Delta t- \frac{\beta x_m}{c}]
 \end{equation}     
\begin{figure}[htbp]
\begin{center}\hspace*{-4.5cm}\mbox{
\epsfysize15.0cm\epsffile{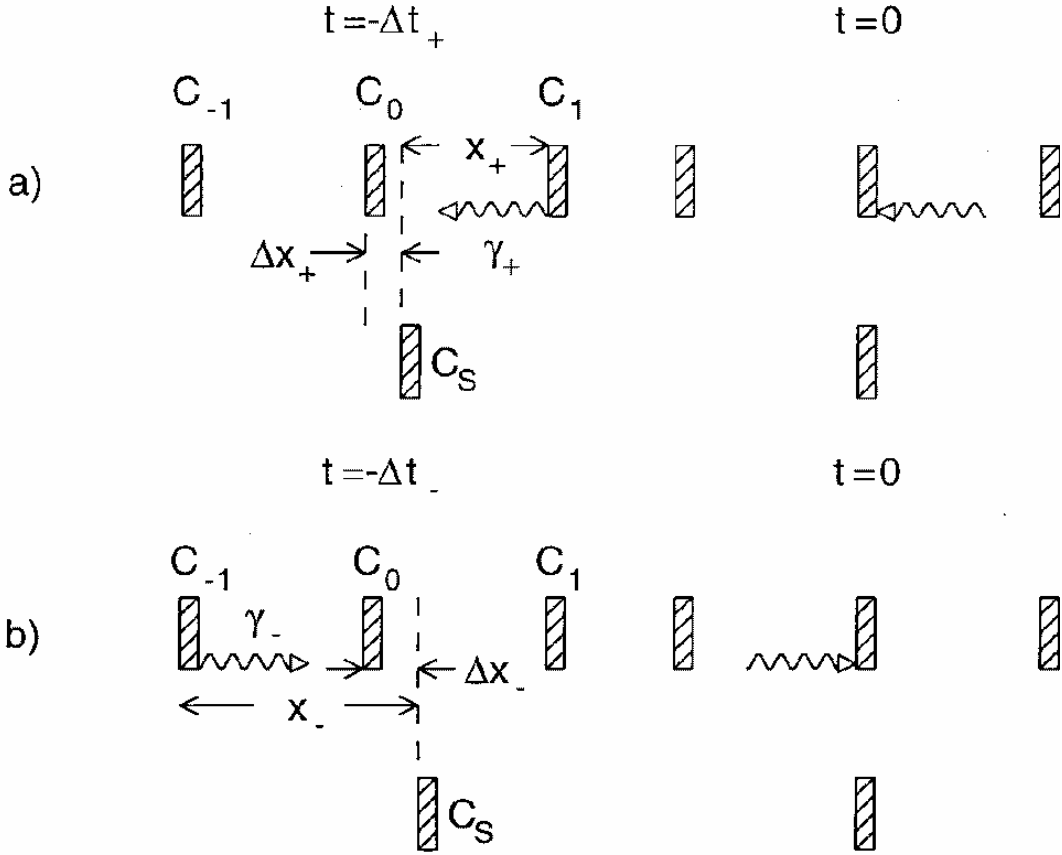}}
\caption{ Propagation time delay effects. In a) the photon $\gamma_+$
 emitted by $C_1$ at time $t=-\Delta t_+$ arrives at the observer beside $C_S$
 at $t=0$. Thus $\Delta t_+ = x_+/c = \Delta x_+/v$. In b) the photon
 $\gamma_-$ emitted by $C_{-1}$ at time $t = -\Delta t_-$ also arrives at $C_S$
 at $t=0$. and $\Delta t_- = x_-/c = \Delta x_-/v$. In a), [b)] the 
 observed clock is receding from [approaching] the observer. Since evidently
 $x_- > x_+$ it follows that $\Delta t_- > \Delta t_+$ so that the effects
 of LPTD are larger for approaching than for receding clocks.
 A corollary (see Ref[5]) is that at $t=0$ the clock $C_{-1}$ appears more distant than
  the clock $C_1$.}
\label{fig-fig11}
\end{center}
 \end{figure}
 There are now two cases to consider:
 \begin{itemize}
 \item[(i)]{$x_m>0$, $C_m$ receding from the observer;}
  \item[(i)]{$x_m<0$, $C_m$ approaching the observer;}     
\end{itemize}
If $|x_m|=c \Delta t$ the observer beside $C_S$ in S will see the time
$t'^D_{m,j}$ indicated by the clock $C_m$ at time $j \tau$. Since $\Delta t$
is, by definition, positive then in case (i) above $x_m$ in (4.25) and (4.26)
is replaced by $c \Delta t$. Eliminating $\Delta t$ between the equations,
after this replacement, gives for the apparent time:
\begin{equation}
 t'^D_{m,j}  = \gamma [ j (1-\beta) - \beta m]\tau~~~(x_m>0)  
\end{equation}
In case (ii) $x_m$ in Eqns.(4.25),(4.26) is replaced by $-c \Delta t$, 
giving the solution:
\begin{equation}
 t'^D_{m,j}  = \gamma [ j (1+\beta) + \beta m]\tau~~~(x_m<0)  
\end{equation}
so that
\begin{eqnarray}
t'^D_{m,j+1}- t'^D_{m,j}& = & \sqrt{\frac{1-\beta}{1+\beta}} \tau~~~(x_m>0) \\
t'^D_{m,j+1}- t'^D_{m,j}& = &\sqrt{\frac{1+\beta}{1-\beta}} \tau~~~(x_m<0)
\end{eqnarray}
Comparing Eqns(4.29),(4.30) with (4.20) it can be seen that, when the LPTD 
are taken into account the TD formula for a local clock is replaced by the 
Relativistic Doppler Effect formulae Eqns(4.29),(4.30). Actually, for $x_m<0$
the clock appears to run fast, not slow. Just, as pointed out in Refs.[2,3,4],
 a moving sphere does not appear flattened by the LFC effect, Eqns.(4.29)
 and (4.30) demonstrate that a moving local clock does not show the TD effect.
 Weinstein~\cite{x5} considered length measurements (for example the distance
 between successive clocks on the train in the present example) under the 
 same conditions as the time measurements decscribed by Eqns.(4.29), (4.30)
 where a single observer is close to a moving object. If $l_0$, $l$ denote
 the lengths of an object viewed in S', S then the relation between $l_0$
 and $l$ is given by the replacements $\tau \rightarrow l_0$, 
 $\Delta t' = l$ in Eqns.(4.29),(4.30). Thus an approaching clock
 (apparently running fast) appears more distant than a receding clock
 which is apparently running slow. Neither the LFC nor the TD effects are
 directly observed when LPTD are taken into account.
 \par It is interesting to note the identity of Eqns.(4.29),(4.30) with 
 the usual Relativistic Doppler Shift formulae which, following Ref.[1] are
 usually derived by considering the LT properties of Electromagnetic Waves.
 Here they have been derived from considerations of space-time geometry,
 as applied to events corresponding to the emission or absorption of
 photons. No use is made of the wave concept. 
 \par Writing Eqns.(4.27),(4.28) in terms of $(n,j)$ gives the equations:
\begin{eqnarray}
t'^D_{n,j}& = &\gamma [j-\beta n] \tau~~~(n>0) \\
t'^D_{n,j}& = &\gamma [j+\beta n] \tau~~~(n<0)
\end{eqnarray}
Both Eqns.(4.31) and (4.32) yield the result:
\begin{equation}
t'^D_{n,j+1}- t'^D_{n,j} = \gamma \tau
\end{equation}   
Thus the TC effect of Eqn(4.23) is unchanged by LPTD corrections
(they must clearly be the same at fixed $n$ or $x$).
The apparent time delay between the clocks on successive wagons is, 
including the effect of LPTD :
\begin{eqnarray}
t'^D_{n+1,j}- t'^D_{n,j}& = &-\gamma \beta \tau~~~(n>0)  \\
t'^D_{n+1,j}- t'^D_{n,j}& = &\gamma \beta \tau~~~~(n<0)
\end{eqnarray}                
Comparing with Eqn.(4.24) it can be seen that the LPTD increases the
absolute size of the delay and, for $n<0$ ( EC approaching the observer )
changes the sign of the effect.
\par The effect of LPTD corrections on the clock $C_m$ may be calculated by 
taking the difference between $t'^D_{m,j}$ given by Eqn.(4.27) or (4.28) and
$t'_{m,j}$ given by Eqn.(4.19). The results are:
\begin{eqnarray}
\Delta'_{j,m} \equiv t'^D_{m,j}- t'_{m,j}& = &\frac{[\gamma^2(1-\beta)-1]}
{\gamma} [m+j]~~~(x_m>0)  \\
\Delta'_{j,m} \equiv t'^D_{m,j}- t'_{m,j}& = &\frac{[\gamma^2(1+\beta)-1]}
{\gamma} [m+j]~~~(x_m<0)
\end{eqnarray} 
\par The apparent times shown by the EC at $t=0$, $t=\tau$, taking into account
LPTD are presented in Table 6 and shown, for the special case $\beta = 0.6$
in Fig 12. Included also in Table 6 and Fig 12 is the `Magic Clock'
$\overline{C}_M$ situated at $x'=x'_M$ where :
\begin{equation}
 x'_M = -\frac{L}{\gamma \beta}\left[1-\sqrt{\frac{1-\beta}{1+\beta}} \right]
\end{equation}
which indicates the same time as $C_S$ at $t = \tau$. Table 6 and Fig.12 show the perhaps
surprising result that the EC situated symmetrically in $x$ relative to $C_S$ at time
$t = 0$ and $t = \tau$ apparently lag $C_S$ by identical times. This is because the time
asymmetry between positive and negative $x$ produced by the LT (see Fig.10) is exactly 
compensated by the LPTD. For example. at $t = \tau$ the LT gives the result for $\beta = 0.6$
 that $C_0$ lags $C_S$ by 0.2$\tau$. However, the longer time delay from $C_{-2}$ as compared 
 to $C_0$ (see Fig.11) means that after correcting for LPTD, $C_{-2}$ appears also to lag 
 $C_S$, by just the same amount as $C_0$, which has a smaller LPTD correction.
\begin{figure}[htbp]
\begin{center}\hspace*{-4.5cm}\mbox{
\epsfysize12.0cm\epsffile{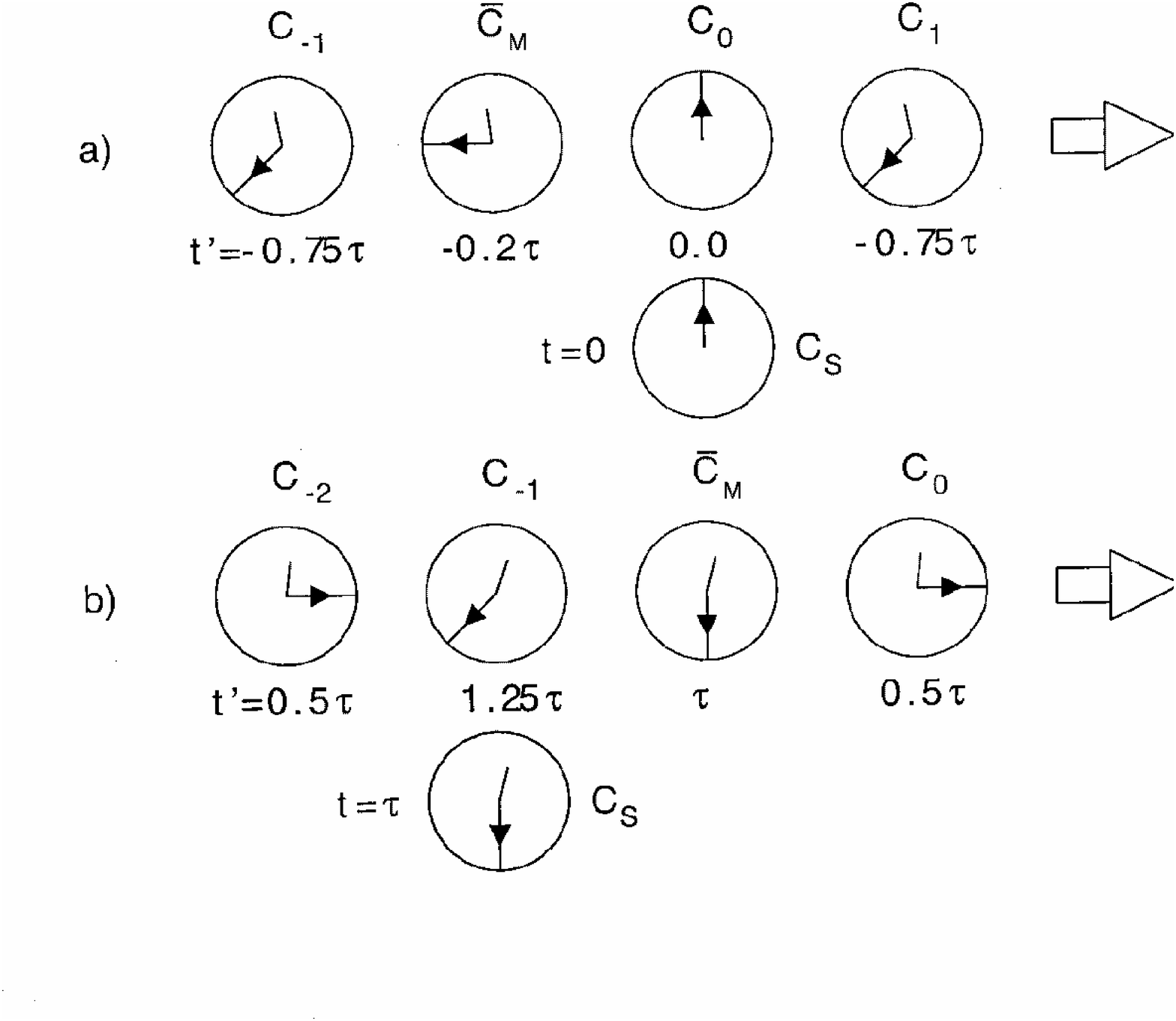}}
\caption{ As Fig.10, but including the effects of LPTD.}
\label{fig-fig12}
\end{center}
 \end{figure}    
\begin{figure}[htbp]
\begin{center}\hspace*{-0.5cm}\mbox{
\epsfysize15.0cm\epsffile{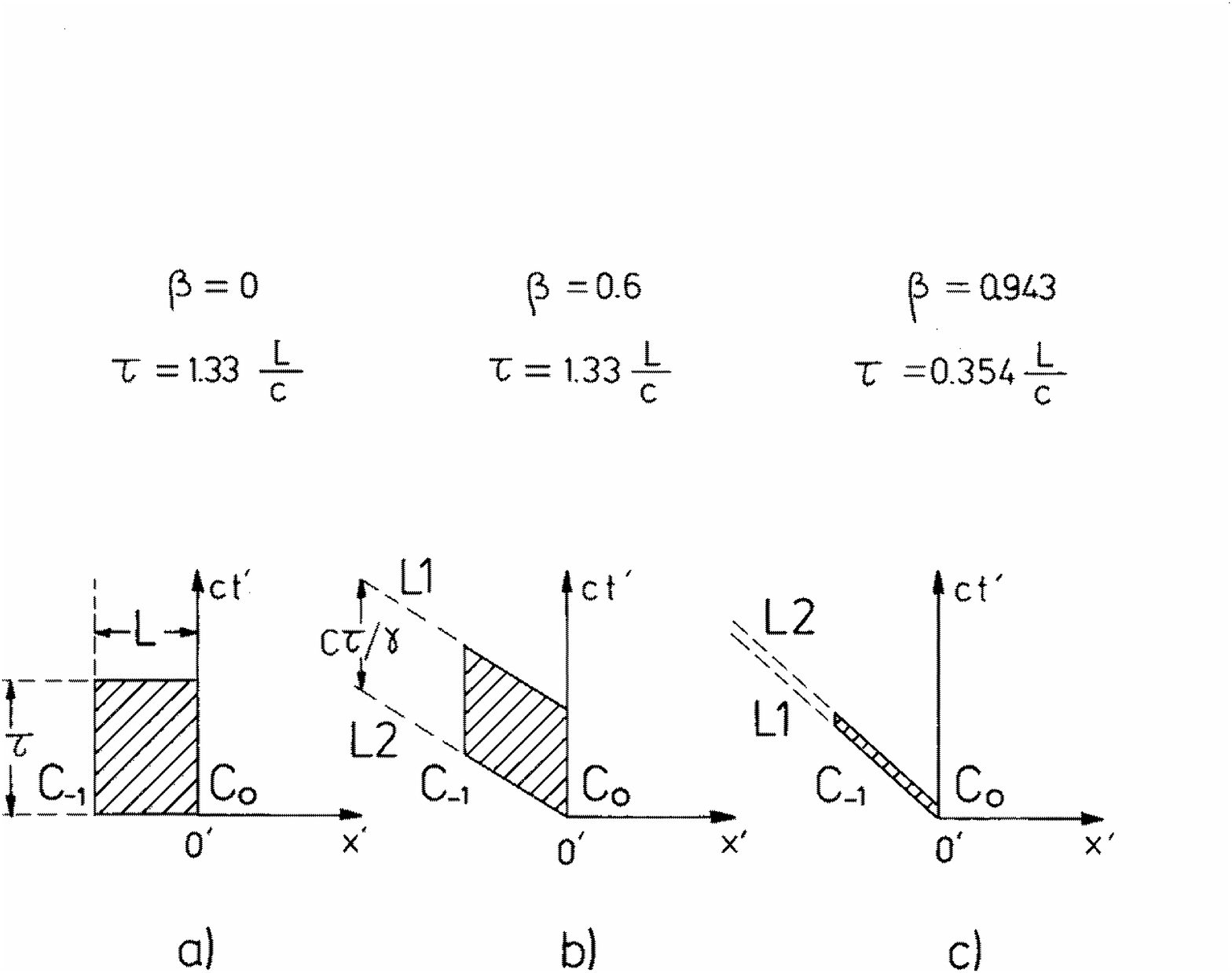}}
\caption{ The domains of ($x'$, $ct'$) space (cross-hatched)
 of the wagon holding the clock $C_0$ (see Fig.9) seen
 by an observer in S' during the time $0<t<\tau$. a),b),c) are for 
 $\beta = 0,0, 0.6, 0.943$ respectively. Without effects of LPTD, as in
 the case
 of an observer at a large transverse distance from the train.}
\label{fig-fig13}
\end{center}
 \end{figure}  
\begin{figure}[htbp]
\begin{center}\hspace*{-0.5cm}\mbox{
\epsfysize15.0cm\epsffile{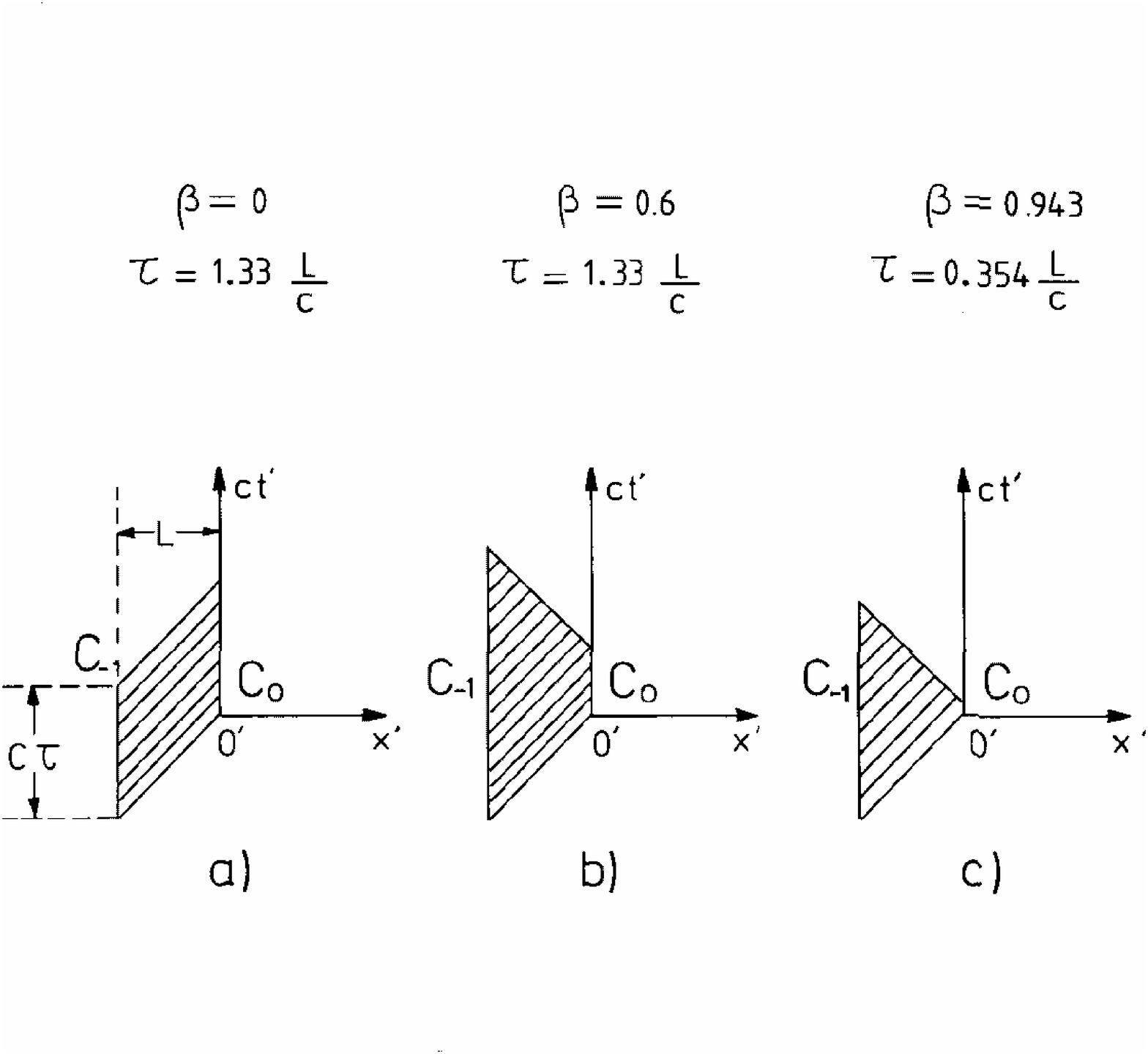}}
\caption{As Fig.13, for an observer close to the train, and including
 the effects of LPTD.}   
\label{fig-fig14}
 \end{center}
\end{figure}   
 \par Finally in this section the region of the space-time domain of S',
  that is visible to the observer in S is discussed. In particular the
  space-time observations which may be made of the wagon holding the EC $C_0$
  during the period $0<t<\tau$ when it is passing by the Standard Clock $C_S$
  will be considered. The situation is shown for the three cases: 
  $\beta = 0.0,~0.6,~0.943$ ($\gamma =1.0,~1.25,~3.0$) in Figs 13a),b),c)
  respectively. For $\beta=0$ the history of each part of the wagon may be 
  observed in an unbiased manner over the whole period. When the wagon is
   moving as in Fig 9b), late times at the front and early times at the 
   rear of the wagon are no longer observable. The observable region of the
   ($x',~ct'$) plane is only that between the lines L1, L2 (Fig 13b)) where:
 \begin{eqnarray}
 L1~:  ct' & = & -\beta x'  \\
 L2~:  ct' & = & -\beta x'+\frac{c \tau}{\gamma}
 \end{eqnarray}
 As $\beta$ approaches one (Fig 13c) the observable domain occupies only a
  narrow region around the backward light cone of the origin of S'. So, although
  the wagons of the train are more concentrated in the field of vision of an
  observer in S, due to the LFC, the fraction of the total space-time area of
  S' that may be observed becomes vanishingly small. Note that the boundaries
  of the observable area in the space-time of S' are easily read off from the
  apparent time of the clocks $C_{-1}$, $C_0$ recorded in Table 5. Eqns.(4.39)
  and (4.40) are derived from Eqn.(4.19) with $j=0,~1$ respectively on 
  making the replacements: $\tau \rightarrow L/\gamma v$,
   $m \rightarrow = x'/L$.  
  The situation shown in Fig 13 Corresponds to the observation of the train
  at a distance such that the angle subtended by the wagon between $C_0$ and 
  $C_{-1}$ at the observer is small. In this case the effects of LPTD
   essentially cancel. It is interesting to compare this with the case of an
   observer close to the train when the LPTD of photons moving almost parallel
   to the train must be taken into account. The ($x'$, $ct'$) domain seen by such
   an observer, for the same conditions as in Fig 13, is shown in Fig 14.
   It is derived in a similar way as for Fig.13, starting from Eqn.(4.28) instead
   of Eqn.(4.19).
  When the train is moving the observable range of $t'$ is always greater at the
 rear end (position of $C_{-1}$) than at the front (position of $C_{0}$) of the wagon.
 As $\beta \rightarrow 1$ the $t'$ range at $C_0$ vanishes and that at $C_{-1}$
  approaches a constant $-L/c < t' < L/c$, corresponding to the full region between
  the forward light cone ($x' = c t'$) and the backward light cone ($x' = -c t'$)
  of the origin of S'.
\SECTION{\bf{Discussion}}
The different space-time effects (apparent distortions of space or time) in 
Special Relativity that have been discussed above are summarised in Table 7.
These are the well-known LFC and TD effects, Space Dilatation (SD) introduced
 in Section 3 above, and Time Contraction (TC) introduced in Section 4. Each 
 effect is an observed difference $\Delta q$ ($q=x,x',t,t'$) of two space or time 
 coordinates ($\Delta q = q_1-q_2$) and corresponds to a constant projection
 $\Delta \tilde{q} = 0$ ($\tilde{q} \ne q$)
 in another of the four variables $x$, $x'$, $t$, $t'$ of the LT. 
 As shown in Table 7, the LFC, SD, TC and TD effects correspond, respectively, to the
 $\Delta t$, $\Delta t'$, $\Delta x$ and $\Delta x'$ projections. After making this
 projection, the four LT equations give two relations among the remaining three
 variables. One of these describes the `space-time distortion' relating 
 $\Delta t'$ and $\Delta t$ or $\Delta x'$ and $\Delta x$ while the other gives
 the equation shown in the last column, (labelled `Complementary Effect') in 
 Table 7. These equations relate either $\Delta x$ to $\Delta t$ (for SD and TD)
 or $\Delta x'$ to $\Delta t'$ (for LFC and TC). It can be seen from the 
 Complementary Effect relations that the two space-time points defining the effect  
 (of space-time distortion) are space-like separated for LFC and SD and time-like
 separated for TC and TD.
 \par For example, for the LFC when $t_1=t_2=t$, the LT equations for the two
 space-time points are:
 \begin{eqnarray}
  x'_1 & = & \gamma (x_1-v t) \\
  x'_2 & = & \gamma (x_2-v t) \\
  t'_1 & = & \gamma (t-\frac{\beta x_1}{c}) \\
  t'_2 & = & \gamma (t-\frac{\beta x_2}{c}) 
  \end{eqnarray}
  Subtracting (5.1) from (5.2) and (5.3) from (5.4) gives:
 \begin{eqnarray}
 \Delta x' & = & \gamma \Delta x \\
 \Delta t' & = & -\frac{\gamma \beta}{c} \Delta x
 \end{eqnarray}
\begin{table}
\begin{center}
\begin{tabular}{|p{1.45in}|c|c|c|c|} \hline
 Name & Observed Quantity & Projection & Effect & Complementary Effect \\
\hline
Lorentz-Fitzgerald Contraction (LFC) & $\Delta x$ & $\Delta t = 0$   
& $\Delta x = \frac{1}{\gamma} \Delta x'$  & $\Delta x' = - \frac{c}{\beta} \Delta t'$ \\
\hline
\mbox{Space Dilatation} (SD) & $\Delta x$ & $\Delta t' = 0$ 
 & $\Delta x = \gamma \Delta x'$  & $\Delta x = \frac{c}{\beta} \Delta t$ \\
\hline 
\mbox{Time Contraction} (TC) & $\Delta t'$ & $\Delta x = 0$   
& $\Delta t' = \gamma \Delta t$  & $\Delta x' = - c \beta \Delta t'$ \\
\hline
\mbox{Time Dilatation} (TD) & $\Delta t'$ & $\Delta x' = 0$   
& $\Delta t' = \frac{1}{\gamma} \Delta t$  & $\Delta x =  c \beta \Delta t$ \\
\hline
\end{tabular}
\caption[]{ The different apparent distortions of space-time in Special Relativity
(see text).  }      
\end{center}
\end{table}
 Eqn.(5.5) describes the LFC effect, while combining Eqns.(5.5) and (5.6) to
 eliminate $\Delta x$ yields the equation for the Complementary Effect.
 By taking other projections the other entries of Table 7 may be calculated in 
 a similar fashion. It is interesting to note that the TD effect can be derived
 directly from the LFC effect by using the symmetry of the LT equations.
 Introducing the notation: $s \equiv ct$, the LT may be written as:
\begin{eqnarray}
 x' & = & \gamma (x-\beta s) \\
 s' & = & \gamma (s-\beta x)
 \end{eqnarray}
 These equations are invariant \footnote{ Actually the transformation $T2$ yields
 the inverse of the LT (5.7),(5.8). The inverse equations may then be solved to
 recover (5.7) and (5.8)} under the following transformations:
\begin{eqnarray} 
 T1~~~& : & x \leftrightarrow s,~~~ x' \leftrightarrow s' \\
 T2~~~& : & x \leftrightarrow x',~~~ s \leftrightarrow s',~~~ \beta
 \rightarrow -\beta     
\end{eqnarray}
Writing out the LFC entries in the first row of Table 7, replacing $t$, $t'$ by
$s/c$, $s'/c$ ; gives  
\[~~~\Delta x~~~~\Delta s = 0~~~~\Delta x = \frac{\Delta x'}{\gamma}~~~~
\Delta x' = -\frac{\Delta s'}{\beta}  \]
Applying $T1$ to each entry in this row results in:
\[~~~\Delta s~~~~\Delta x = 0~~~~\Delta s = \frac{\Delta s'}{\gamma}~~~~
\Delta s' = -\frac{\Delta x'}{\beta}  \]
Applying $T2$:
\[~~~\Delta s'~~~~\Delta x' = 0~~~~\Delta s' = \frac{\Delta s}{\gamma}~~~~
\Delta s = \frac{\Delta x}{\beta}  \]
Replacing $\Delta s$, $\Delta s'$ by $c\Delta t$, $c\Delta t'$
 yields the last row of Table 7 which describes
the TD effect. Similarly TC can be derived from SD (or vice versa) by successively
applying the transformations $T1$, $T2$.
\par A remark on the `Observed Quantities' in Table 7. For the LFC, SD the observed
 quantity is a length interval in the frame S. The apparent space distortion occurs
 because this length differs from the result of of a similar measurement made on
 the same object in its own rest frame. $\Delta x'$ is not directly measured
 at the time of observation of the LFC or SD. It is otherwise with the time
 measurements TD, TC. Here the time interval {\it in their own rest frame}
 indicated by a moving clock (TD), or 
 different equivalent clocks at the same position in S (TC),  
 is supposed to be directly observed and compared
 with the time interval $\Delta t$ registered by an unmoving clock in the
  observer's rest frame.
 Thus the effect refers to two simultaneous observations by {\it the same observer}
 not to separate observations by {\it two different observers} as in the case 
 of the LFC and SD.
 \par Einstein's great achievement in his first paper on Special Relativity~\cite{x1}
 was, for the first time, to clearly disentangle in Classical Electromagnetism, the 
 purely geometrical and kinematical effects embodied in the Lorentz Transformation
 from dynamics. In spite of this, papers still appear from time to time in the
 literature claiming that moving objects `really' contract~\cite{x6} or that 
 moving clocks `really' run slow~\cite{x7} for dynamical reasons, or even that such
 dynamical effects are the true basis of Special Relativity and should be 
 taught as such~\cite{x8}. As it has been shown above that a moving object can
 apparently shrink or expand, and identical moving clocks can apparently run fast
 or slow, depending only on how they are observed, it is clear that they cannot
 `really' shrink, or run slow, respectively. If a moving object actually shrinks for 
 dynamical reasons it is hard to see how the same object, viewed in a different
 way (in fact only illuminated differently in its own rest frame) can be seen to
 expand. Certainly both effects cannot be dynamically explained. In fact the 
 Lorentz Transformation, as applied to space-time, describes only the 
 {\it appearance} of space-time events, a purely geometrical property. The 
  apparent distortions are of geometrical origin, the space-time analogues of
  the apparent distortions of objects in space, described by
  the laws of perspective, when they are linearly projected into a two dimensional
  sub-space by a camera or the human eye. 
 \par In conclusion the essential characteristics of the two `new' space-time 
 distortions discussed above are summarised :
 \begin{itemize} 
 \item \underline{Space Dilatation (SD):} If a luminous object, lying along the Ox'
 axis at rest in the frame S', has a short luminous lifetime in this frame,
 it will be observed from
 a frame S, in uniform motion relative to S' parallel to Ox' at the velocity -$\beta c$,
 as a narrow line, perpendicular to the x-axis, moving with the velocity
 $c/\beta$ in the same direction as the moving object. The total distance swept out
 along the $x$-axis 
 by the moving line during the time $\beta l_0/(c \sqrt{1-\beta^2})$, for which
 the moving line image exists, is 
 $l_0/\sqrt{1-\beta^2}$ where $l_0$ is the length of the object as observed in S'.
 Thus the apparent length of the object when viewed with a time resolution
 much larger than $\beta l_0/(c \sqrt{1-\beta^2})$ is  $l_0/\sqrt{1-\beta^2}$.
 Any effects of LPTD are not taken into account.
 \item \underline{Time Contraction (TC): } The equivalent clocks in the moving frame 
 S', viewed at the same position in the stationary frame S, apparently run faster by a factor
 $1/\sqrt{1-\beta^2}$ relative to a clock at rest in S.
\end{itemize}
 {\bf Acknowledgements} 
\par I thank G.Barbier and C.Laignel for their valuable help in the preparation
 of the figures. 
\pagebreak
 
\end{document}